\documentclass[10pt]{article}

\usepackage[pdfpagemode=UseNone,pdfstartview=FitH,breaklinks,backref=page]{hyperref}

\usepackage{amsmath,amsthm,amssymb,amsfonts,latexsym,graphicx,enumitem,booktabs}
\usepackage[numbers]{natbib}

\usepackage{mathtools}

\usepackage{mathcomp}

\usepackage{array}
\newcommand{\PreserveBackslash}[1]{\let\temp=\\#1\let\\=\temp}
\newcolumntype{C}[1]{>{\PreserveBackslash\centering}p{#1}}
\newcolumntype{R}[1]{>{\PreserveBackslash\raggedleft}p{#1}}
\newcolumntype{L}[1]{>{\PreserveBackslash\raggedright}p{#1}}

\usepackage[T2A]{fontenc}

\usepackage[russian,english]{babel}
\usepackage{hyphenat}
\makeatletter
\renewcommand\tableofcontents{%
    \@starttoc{toc}%
}
\makeatother

\allowdisplaybreaks[4]

\usepackage[toc,page,title,titletoc]{appendix}

\theoremstyle{definition}

\newlength{\IndentI}
\newlength{\IndentII}
\newlength{\IndentIII}
\newlength{\IndentIV}
\title{On the nineteenth-century origins of significance testing and p-hacking}
\date{\today}

\author{Glenn Shafer, Rutgers University}

\begin{document}

\maketitle

\begin{abstract}
Although the names \emph{significance test}, \emph{p-value}, and \emph{confidence interval} came into use only in the 20th century, the methods they name were already used and abused in the 19th century.  Knowledge of this earlier history can help us evaluate some of the ideas for improving statistical testing and estimation currently being discussed.

This article recounts first the development of statistical testing and estimation after Laplace's discovery of the central limit theorem and then the subsequent transmission of these ideas into the English-language culture of mathematical statistics in the early 20th century.  I argue that the earlier history casts doubt on the efficacy of many of the competing proposals for improving on significance tests and p-values and for forestalling abuses.  Rather than further complicate the way we now teach statistics, we should leave aside most of the 20th-century embellishments and emphasize exploratory data analysis and the idea of testing probabilities by betting against them.
\vspace*{0.4in}
\end{abstract}

\tableofcontents

\section{Introduction}

At a time when the evaluation of statistical evidence has become ever more important, standard methods for this evaluation are under increasing scrutiny.  Some statisticians argue against  the continued use of the phrase \emph{statistical significance}; others deplore the way \emph{p-values} are taught and used.   Numerous competing proposals for improving the presentation of statistical results are being advanced.  For some observers, the entire enterprise of objective judgment based on statistics is threatened by the practice of searching for statistical significance, \emph{p-hacking} as it is now called.

Those addressing this problem draw on experience from many domains of application.  History may also have a role to play.  Marie-France Bru and Bernard Bru have made the case for history's role with these eloquent words:
\begin{quote}
To penetrate to the reasons of things, look at how they have gradually been revealed in the course of time, in their progression and in their ruptures, if any.%
\footnote{Translated from \cite[pp. 301--302]{Bru/Bru:2018}:  Pour p\'en\'etrer les raisons des choses, voyons comment elles se sont d\'evoil\'ees progressivement au cours du temps dans leurs encha\^inements et leurs ruptures, s'il s'en trouve.}
\end{quote}
By studying the historical record, we may gain more understanding of how the space of possibilities in which we are struggling was constructed and why it has persisted through earlier crises.  We may even find discarded paths of thought that might merit renewal.

Some authors have traced current difficulties with significance testing back to the 20th-century statisticians R.\ A.\ Fisher and Jerzy Neyman.  The development of Fisher's and Neyman's ideas and the conflicts between them form a fascinating and important chapter of intellectual history,%
\footnote{See especially \cite{Lehmann:2011}.}
and more attention to that history might diminish errors and abuses resulting from university teaching that fails to distinguish clearly between Fisher's ideas about evidence and Neyman's ideas about decision-making.%
\footnote{See for example \cite{Gigerenzer:2018}.}
Debates about statistical inference since Fisher and Neyman, concerning objective and subjective probability and the roles of Bayes's rule and likelihood, are also important and in some cases venerable enough to qualify for historical as well as philosophical study.%
\footnote{See \cite{Mayo:2018} and its many references.}
For a full picture, however, we also need to reach further back in history.  For in many respects, Fisher's and Neyman's ideas had already been developed in the 19th-century, in traditions launched by Laplace and Gauss.  

Nineteenth-century parallels to what has happened in the 20th century and the first two decades of the 21st give us reason to question some diagnoses for current difficulties.  Abuses being decried today---searching for significant results, unsupported modeling assumptions for observational data, etc.---arose in the 19th century as well.  These abuses cannot be blamed on the terms \emph{statistically significant} and \emph{p-value}, which were not in use in the 19th century.  It also becomes more difficult to hope that an emphasis on confidence intervals can diminish abuses of significance testing when we learn that 19th-century p-hacking grew out of the teaching of confidence intervals.

This history also shows us rupture.  The Laplacean tradition was heavily scarred by 19th-century abuses, to the extent that it was ridiculed and largely forgotten, especially in France.  Current neglect of the 19th-century history is due in part to the fact that R.\ A. Fisher and many of his successors felt that they were building on ruins. As Fisher's daughter Joan Fisher Box said this bluntly in 1978 \cite[p. 62]{Box:1978}:
\begin{quote}
The whole field was like an unexplored archeological site, its structure hardly perceptible above the accretions of rubble.
\end{quote}
This enduring influence of this view is illustrated by David J.\ Hand's approving citation of Box's words in 2015 \cite[p. 2]{Hand:2015}.

The historical account in this paper begins, in Section \ref{sec:19th}, by discussing Laplace's large-sample statistical theory in the 19th century---how it was created, how it differed from Gauss's theory of least squares, how it was popularized, and how it was discredited.  Laplace's theory was based on the central limit theorem, which he discovered in 1810 and vigorously promoted until his death in 1827.  From this theorem, using the method of least squares but without knowing probabilities for measurement errors or prior probabilities, we can obtain large-sample confidence limits for unknown quantities and hence significance tests for whether the quantities are zero.  Gaussian least squares, which emphasized finding the best estimates rather than practical certainty for limits on errors, came to dominate work in astronomy and geodesy, but Laplace's large-sample theory was widely used in the human sciences once Joseph Fourier and Sim\'eon-Denis Poisson made it accessible to statisticians.  The uses included p-hacking and inferences based on questionable assumptions.  

Although the misuse of Laplace's theory so discredited it in France that it was practically forgotten there by the end of the 19th century, it was still taught and used elsewhere.  Section~\ref{sec:later} sketches its transmission into Britain and the United States: how limits for practical certainty were variously expressed in terms of probable errors, moduli, standard errors, and finally, in Fisher's \emph{Statistical Methods}, tail probabilities, and how the terms \emph{significance} and  \emph{p-value} emerged.  The use of \emph{significant} as a technical term in statistics derives from its use by Francis Edgeworth in 1885, and some of the confusion associated with the word results from Edgeworth and Karl Pearson using it in a way that is no longer readily understood.  The term \emph{p-value} appears in statistics in the 1920s, deriving from the much older ``value of P''.

Section~\ref{sec:conclusion} looks at lessons we might draw from this history.  One lesson is that p-hacking and other abuses of statistical testing need not be blamed on 20th-century innovations; the same abuses arose already in the 19th century.  The difficulties go deeper, and remedies must go deeper.  I argue that we should simplify, not complicate, our teaching of error limits, that we should acknowledge and even emphasize pitfalls, and that we should teach statistical testing in betting terms, reviving a fundamental aspect of probabilistic reasoning that has been progressively minimized for three centuries.

Appendix~\ref{app:FC} documents the history set out here with quotations from 19th- and early 20th-century authors.  Those quoted are only a few of the many who wrote on statistical estimation and testing during this period, and only a few of their words are quoted, but these words may help readers judge for themselves how the logic and the pitfalls of statistical testing did and did not change over the period.  

Appendix~\ref{app:bernoullian} quotes some authors who have distinguished between \emph{Bernoullian} and \emph{Bayesian} statistical methods.

\section{Laplace's theorem}\label{sec:19th}

The sum of a large number of independent random variables is approximately normal.  This theorem, with any of various regularity conditions that would make it true, is now called the \emph{central limit theorem}, but there is justice in calling it \emph{Laplace's theorem}.  Pierre Simon Laplace proved it in 1810, with his characteristic neglect of regularity conditions, and fully recognized its importance.  It was named the central limit theorem (\emph{zentraler Grenzwertsatz der Wahrscheinliehkeitsrechnung}) by Georg P\'olya in 1920.

\subsection{Laplace's discovery}

The integral of $e^{-t^2}$ appeared in probability theory beginning in 1733, with Abraham De Moivre's asymptotic approximation for the sum of an interval of binomial probabilities.  But the notion of a probability distribution with a density of the form 
\begin{equation}\label{eq:gade}
      f(y) =  \frac{h}{\sqrt{\pi}} e^{-h^2 y^2}
\end{equation}
appeared only in 1809, when Carl Friedrich Gauss advanced it as a hypothetical distribution for errors of measurement, awkwardly justifying it by the fact that it makes the arithmetic average of the measurements the mode of the posterior probabilities for the quantity measured.  Perhaps partly inspired by Gauss's insight, but certainly also inspired by Fourier's emerging theory of the heat equation, Laplace soon afterwards arrived at his theorem, which gave~\eqref{eq:gade} as the distribution of the arithmetic average of a large number of independent variables, regardless of their individual distribution.%
\footnote{Many authors have detailed the interplay between Laplace and Gauss: \cite{Bru/Bru:2018,Farebrother:1999,Fischer:2011,Gorroochurn:2016,Hald:1998,Sheynin:1977a,Sheynin:1979,Sheynin:1994,Stigler:1986}.}

Laplace first applied his theorem to a problem that had long concerned him, that of testing the hypothesis that planetary orbits were distributed randomly.  But he quickly realized that he could also use it to justify estimation by least squares.  This inspired both his monumental \emph{Th\'eorie analytique des probabilit\'es} (1812) and his more verbal \emph{Essai philosophique sur les probabilit\'es} (1814).  He also wrote to colleagues throughout Europe to explain the importance of his theorem, illustrating its power with examples.%
\footnote{The scale of this correspondence has only recently become known, with Roger Hahn's publication of Laplace's surviving correspondence by \cite{Hahn:2013}; for a summary, see \cite[vol.~2, pp.~455ff]{Bru/Bru:2018}.}

\subsection{Direct and inverse probability}

In the 1770s and 1780s, Laplace had developed Bayes's rule, which Thomas Bayes and his friend Richard Price had formulated only for the elementary case of independent trials of an event with constant probability.  In the 1780s, Laplace had tried to base a theory of errors on his generalization of Bayes's rule, but he had been stymied by an inability to calculate the distribution of averages or other functions of more than a few observations.

Laplace's 1810 theorem solved this problem of calculation in a spectacular way.  Not only could Laplace now calculate probabilities for the average of many independent quantities; he could do so without even knowing probabilities for the individual quantities.  Bayes's rule also now seemed less interesting.  The very concentrated normal distribution of the average would dominate any prior probabilities, so that Bayes's rule would give the same result as a direct argument in the style of Jacob Bernoulli, like the arguments Thomas Simpson and Daniel Bernoulli had earlier proposed for a theory of errors.%
\footnote{This was explained very clearly by Cournot in the binomial case; see~\S\ref{subsec:Cournot}.  Bienaym\'e also discussed aspects of this picture; see \cite[p.~102]{Heyde/Seneta:1977}.}
Laplace did not disown Bayes's rule, but he de-emphasized it in his \emph{Th\'eorie analytique}, and in the applications of his theorem he usually just stated the Bernoullian argument \cite{Hald:1998,Bru/Bru:2018}.  This inattention to the difference between Bayesian and Bernoullian arguments continued in the Laplacean tradition throughout the 19th century and into the time of Karl Pearson.  It allowed mathematical statisticians to communicate with each other easily, regardless of whether (like Antoine Augustin Cournot) they rejected Bayesian arguments or whether (like Edgeworth and Pearson) they saw a Bernoullian argument as a shortcut to a Bayesian answer with roughly uniform prior probabilities.

Laplace did not give names to the two modes of argument that I am calling Bernoullian and Bayesian.  In 1838 the British mathematician Augustus De Morgan called them \emph{direct probability} and \emph{inverse probability}, respectively \cite{De-Morgan:1838,Walker:1929},%
\footnote{Fourier had apparently used the corresponding French terms earlier in his teaching \cite{Dale:1999}.}   
and these terms were widely used in English for over a century.  Since the 1970s, the two modes have been called \emph{frequentist} and \emph{Bayesian}.  Because \emph{frequentism} also names a view about the meaning of probability, clarity may be served if, as I have done here, we use \emph{Bernoullian} instead of \emph{frequentist} for the non-Bayesian mode of argument.\label{p:bernoullian}%
\footnote{Authors who have used \emph{Bernoullian} in this way include Edgeworth \cite{Edgeworth:1918}, Richard von Mises \cite[p.~5]{vonMises:1919}, A.\ P. Dempster \cite{Dempster:1966}, and Ian Hacking \cite{Hacking:1990}.  See Appendix~\ref{app:bernoullian}.}

The 19th-century Bernoullian arguments produced what we now call \emph{confidence intervals} and \emph{two-sided significance tests}.  The widespread belief that these concepts were invented only in the 20th century may be due in part to some authors' reluctance to recognize the merit and coherence of statistical theory not based on 20th-century foundations for probability theory; see \S\ref{subsec:Neyman}.

\subsection{Laplacean and Gaussian least squares}\label{subsec:LG}

Gauss appreciated Laplace's theorem, but he and those who continued his work on least squares tempered their interest in it with a concern about systematic errors, computational accuracy and efficiency, and other practical problems.  Moreover, Gauss eventually gave an alternative justification of least squares that applies when samples are small.  This we now call the \emph{Gauss-Markov theorem}:  least-squares estimators have the least variance among unbiased linear estimators when individual errors are unbiased and independent.  

The result was two interacting but distinct schools of thought, one Gaussian, the other Laplacean.  The Gaussian school soon dominated astronomy and geodesy.%
\footnote{The story of the triumph of the Gaussian theory in geodesy has been told in an enlightening dissertation by Marie-Fran\c{c}oise Jozeau \cite{Jozeau:1997}.} 
The Laplacean school, which sought practical certainty from large samples, continued to find adherents in the human sciences.

We can gain insight into how the two schools differed by looking at two tables of tail probabilities for the normal distribution, one published in 1816 by Gauss's disciple Friedrich Wilhelm Bessel, the second published in 1826 by Joseph Fourier, in an exposition of Laplace's theory.  Although Christian Kramp had published a table of values of the integral of $e^{-t^2}$ in 1799 \cite{Kramp:1799}, Bessel's and Fourier's tables appear to be the first published tables of the normal distribution.%
\footnote{In 1783, in the course of a Bayesian analysis of Bernoulli's binomial problem, Laplace gave a method for calculating values of the incomplete integral of $e^{-t^2}$ and mentioned that a table of these values would be useful \cite[p.~79]{Gorroochurn:2016}.  Kramp's table gave values of $\int_\tau^\infty e^{-t^2}dt$ for $\tau$ from $0.00$ to $3.00$, in intervals of $0.01$.  Kramp calculated his table to facilitate the study of refraction, not to facilitate the calculation of probabilities, and because $\int_{-\infty}^\infty e^{-t^2}dt=\sqrt{\pi}$, the entries in his table are not probabilities.  But we obtain probabilities simply by dividing by $\sqrt{\pi}$.  Bessel used Kramp's table to calculate his.}

Bessel's table appears in an article on the orbit of Olber's comet, in a passage translated in \S\ref{subsec:Bessel}.  In this passage Bessel explained why the different equations of condition in a least-squares computation should be weighted differently.  To this end, he introduced the \emph{probable error} of a continuous variable, which had not been previously defined and used.   As the reader may recall, this is the number $r$ such that $P(|X-\mu|<r)=P(|X-\mu|>r)$, where $X$ is the variable and $\mu$ is $X$'s mean.  When $X$ is normal, $r\approx0.6745\sigma$, where $\sigma$ is the standard deviation.  For seven different values of $\alpha$, Bessel's table gives the odds that a normally distributed variable falls more than $\alpha$ probable errors from its mean:
\[
\begin{matrix*}[l]
\alpha   =1 &\cdots\cdots&1:1\\
\alpha   =1.25 &\cdots\cdots&1:1.505\\
\alpha   =1.5 &\cdots\cdots&1:2.209\\
\alpha   =1.75 &\cdots\cdots&1:3.204\\
\alpha   =2 &\cdots\cdots&1:4.638\\
\alpha   =3 &\cdots\cdots&1:30.51\\
\alpha   =4 &\cdots\cdots&1:142.36
\end{matrix*}
\]
The probability of the variable being more than $4$ probable errors from its mean, for example, is approximately $1/(1+142.36)$ or $0.007$.

Fourier's table appeared in a memoir on the use of probability that he included in the 1826 report of the statistics bureau of Paris and its surrounding region. Instead of the probable error, Fourier used as his measure of dispersion the quantity $\sqrt{2}\sigma$, which I will call the \emph{modulus}, following later authors.  The modulus is a natural choice because the density for a normal variable with mean $0$ and modulus $1$ is proportional to $e^{-t^2}$.  One modulus is approximately two probable errors.  For each of $5$ small probabilities (we might call them \emph{significance levels} today), Fourier gave the number $\partial$ such that a normal random variable will be more than $\partial$ moduli from its mean with that probability.
\begin{center}
\def\arraystretch{1.2}
\begin{tabular}{cc}
     $\partial$ &  P\\
     0.47708   & $\frac12$ \\
     1.38591   & $\frac{1}{20}$ \\
     1.98495   & $\frac{1}{200}$ \\
     2.46130   & $\frac{1}{2000}$ \\
     2.86783   & $\frac{1}{20000}$ 
\end{tabular}
\end{center}
There is only a probability of $1$ in 20,000, for example, that a normal variable will be more than about $2.86783$ moduli (or about $4.0557$ standard deviations) from its mean.

Fourier's table differs from Bessel's in two important ways.  First, it uses round numbers for the probabilities rather than for the distance from the mean.  It gives the distance from the mean corresponding to a significance level the reader might have in mind.  Second, it includes much more extreme values.  Whereas Bessel's table extends to only 4 probable errors, Fourier's extends to 2.87 moduli, equivalent to about 6 probable errors and corresponding to a probability two orders of magnitude smaller.  Fourier was interested in identifying limits within which we can be practically certain the deviation from the mean will fall.

Fourier decided that 3 moduli is enough for practical certainty.  His example was followed by a number of other 19th-century authors.

\subsection{Seeing p-hacking in France}

The word ``statistics'' (\emph{Statistik} in German and \emph{Statistique} in French) was coined to refer to information monarchs might want to know about their kingdoms' population and wealth.  The theory of errors, conceived as a tool for astronomy and geodesy, did not fall under this rubric at the beginning of the 19th century.  But Fourier, in his reports for the Paris statistics bureau, applied the Laplacean theory of errors to statistics.  In his 1829 report, for example, he gave estimates and error limits for the average age, in Paris during the 18th century, of men and women when they married and when their first son was born \cite[Table 64]{Fourier:1829}.

Such applications soon made what we now call significance testing popular.%
\footnote{As we know, such tests were already being published and debated in the 18th century.  The first example usually cited is John Arbuthnot's argument, from boy births outnumbering girl births in London 82 years running, that the ratio must be governed by divine providence rather than chance.  A good account of this and subsequent significant tests by Daniel Bernoulli, John Michell, and Laplace is provided by Anders Hald \cite{Hald:1998}.  Barry Gower has provided more philosophical reviews \cite{Gower:1982,Gower:1987}.}
In 1824, Sim\'eon-D\'enis Poisson, who became the leading expert on Laplace's theory after Laplace's death in 1827 and Fourier's in 1830, published a note observing that the ratio of boys to girls was smaller for illegitimate births than for legitimate births \cite{Poisson:1824}.  In 1830, he applied Laplace's theory to decide whether this and other variations in the ratio of boys to girls could have happened by chance \cite{Poisson:1830}.  He concluded that the difference was real, and he also found that the ratio was smaller in Paris than in the rest of France, for both legitimate and illegitimate births.

Were Fourier's and Poisson's arguments valid?  Were the 505 men and 486 women for whom Fourier was able to find the needed data a random sample?  In what sense were the approximately ten million births in France in the decade from 1817 to 1826, which Poisson studied, a random sample from a larger population?  Some French statisticians thought Fourier's and Poisson's calculations were ridiculous.  Among them was André-Michel Guerry, the statistician whose brilliant study of crime in France was commissioned by the Academy of Sciences in 1833 \cite{Guerry:1833,Friendly:2007}.

Cournot, twenty years younger than Poisson, was himself a proponent and brilliant expositor of the Bernoullian version of Laplace's theory, but he perceived another problem with its application to the census by Fourier, Poisson, and statisticians who had imitated them.  The problem is what we now call \emph{p-hacking}.  In 1843, safely after Poisson's death, Cournot published his own book on probability, \emph{Exposition de la th\'eorie des chances et des probabilit\'es} \cite{Cournot:1843}.  In a passage reproduced in translation in \S\ref{subsec:Cournot}, Cournot observed that statisticians had been searching for differences in the sex ratio for all sorts of ways of dividing the population:  legitimate and illegitimate, by season, by birth order, etc.  As the public could not see the extent of the search, they could not evaluate whether a particular apparently remarkable difference might arise by chance.

\subsection{The disappearance of Laplace's theorem in France}

Uncertainty measured by a p-value is often the least of our uncertainties when we are working with data.  One of Laplace's favorite examples of the power of his theorem was his estimation of Jupiter's mass relative to the Sun.  Combining all relevant measurements made by that time, he announced bounds on this ratio, bounds on which he claimed one could bet a million to one.  Five years after Laplace's death, the British astronomer George Biddell Airy showed that the true ratio lay well outside these bounds.  Laplace's supreme confidence, whether in his model, his data, or his calculations, had been misplaced \cite[p.~492]{Bru/Bru:2018}.

This was only one reason for the discredit into which Laplace's theory fell.  Though a champion of Laplace's theorem, Cournot ridiculed Laplace's Bayesian argument, emphasizing that for many questions it is only possible to justify non-numerical ``philosophical probabilities'' \cite{Cournot:1843}.   Cournot's friend Jules-Ir\'en\'ee Bienaym\'e further developed Laplace's theory but spent most of his energies combating faulty applications \cite[notes 27, 29, 62]{Bru/etal:1997},\cite{Bienayme:1838,Heyde/Seneta:1977}.  The nineteenth century saw an unprecedented flood of data, and many of its collectors and users concluded that it could speak for itself; probability was not needed \cite{Hacking:1990}.  By the middle of the century, geodesy, a field dominated by the French before and during Laplace's heyday, had abandoned Laplace's methods, turning instead to the methods developed by Gauss and his followers \cite{Jozeau:1997}.  Mathematicians and philosophers found many other problems in the Laplacean theory \cite{Kamlah:1987,Schneider:1987}.   By the end of the century, the most prominent mathematician in France, Joseph Bertrand, would ridicule Laplace's entire undertaking as a delusion \cite{Bertrand:1889}.  Its disappearance from French mathematics was so thorough that the leading French mathematicians who worked on the central limit theorem in the early 20th century, Borel, Fr\'echet, and L\'evy, were unaware that Laplace had first proven the theorem until this was brought to their attention by foreign colleagues.

\section{Practical certainty}\label{sec:later}

Now we look at how Laplace's theory and Fourier's criterion for practical certainty evolved in the 19th and into the 20th century.  Throughout this period, authors on the theory of errors, practically without exception, can be classified as either Gaussian or Laplacean.  Both schools taught the use of least squares to obtain estimates and used the normal distribution to compute probabilities of error.  But the Gaussian authors, considering their models and Laplace's asymptotics too approximate for extreme conclusions, did not talk about practical certainty, whereas the Laplacean authors usually tried to specify, in one way or another, an interval around the least-squares estimate that is practically certain to include the true value of the quantity being estimated.

Today we measure a variable's distance from its mean in terms of its \emph{standard deviation}, and we sometimes call the standard deviation of an estimator its \emph{standard error}.  But these English terms appeared only at the end of the 19th century;  Karl Pearson introduced \emph{standard deviation} in 1894 \cite{Pearson:1894}, and George Udny Yule introduced \emph{standard error} in 1897 \cite{Yule:1897}.  Earlier writers had other names for the standard deviation, but they more often used the modulus or the probable error; see Table~\ref{ta:airy}.  The probable error was still widely used in the first decades of the 20th century.%
\footnote{Helen Walker's history, written in 1929 \cite{Walker:1929}, is still a good reference for how various authors used and named the various measures.} 

\begin{table}
\renewcommand{\arraystretch}{1.5}
\caption{Geroge Biddell Airy's table relating different measures of dispersion for the normal distribution \cite[p.~24]{Airy:1861}.  Airy's \emph{error of mean square} is the standard deviation.  His \emph{mean error} is the mean of the absolute deviation.  
}\label{ta:airy}
\begin{center}
\begin{tabular}{|l|c|c|c|c|}
\hline
             & 
             Modulus. & 
             \begin{minipage}{1.5cm}\begin{center}Mean\\Error.\end{center}\end{minipage} & 
             \begin{minipage}{1.5cm}\begin{center}\vspace*{0.1cm}Error of\\Mean\\Square.\vspace*{0.1cm}\end{center}\end{minipage} & 
             \begin{minipage}{1.5cm}\begin{center}Probable\\Error.\end{center}\end{minipage} \\
 \hline
 \begin{minipage}{3.5cm}\vspace*{0.1cm}In terms of Modulus\vspace*{0.1cm}\end{minipage} & 1.000000 &  0.564189 & 0.707107 & 0:476948\\
 \hline
 \begin{minipage}{3.5cm}\vspace*{0.1cm}In terms of Mean Error\vspace*{0.1cm}\end{minipage} & 1.7724.54 & 1.000000 & 1.253314 & 0.845369\\
 \hline
 \begin{minipage}{3.5cm}\vspace*{0.1cm}In terms of Error of \\ \hspace*{0.5cm}Mean Square\vspace*{0.1cm}\end{minipage} & 1.414214 & 0.797885 & 1.000000 & 0.674506    \\
 \hline
 \begin{minipage}{3.5cm}\vspace*{0.1cm}In terms of Probable \\ \hspace*{0.5cm}Error\vspace*{0.1cm}\end{minipage} & 2.096665 & 1.182916 & 1.482567 & 1.000000  \\
 \hline
\end{tabular}
\end{center}
\end{table}

\subsection{\emph{La limite de l'\'ecart}}

Equating very high probability with moral certainty is an ancient idea.  The 16th-century Jesuit Luis Molina even applied it to games of chance \cite[pp.~402--403]{Knebel:2000}.  But Molina and his fellow scholastics did not gauge degrees of probability numerically.  It was Jacob Bernoulli's \emph{Ars conjectandi}, published in 1713, that brought moral certainty into the context of a mathematical theory of probability modeled after calculations of chances for dice.  Bernoulli did not settle on a particular level of probability that would suffice for moral certainty; he thought 0.99 or 0.999 might do but suggested that the level be set by magistrates.%
\footnote{For Molina and Bernoulli, a \emph{morally certain} thing was one sufficiently probable that it should, by custom or some other norm, be treated as fully certain.  For the purposes of this paper we may equate their concept of moral certainty with the later concept of \emph{practical certainty}.}
So far as I know, Laplace also never specified a level of probability that would suffice for certainty.  Fourier may have been the first to do so.  As we have seen, Fourier considered a statement certain if the probability of its being wrong is only 1 in 20,000, and for an interval based on Laplace's theorem, this corresponds to 2.87 moduli, a number that Fourier rounded to 3.

Later authors sometimes set exigent standards for certainty in theoretical discussions but then relaxed them in applications.  Sim\'eon-Denis Poisson, in his 1837 book on probability, first mentioned 4 or 5 moduli but relaxed this to 2 moduli when he turned to examples, even writing at one point that the probability of an estimate being within 2 moduli, 0.9953, is very close to certainty; see \S\ref{subsec:Poisson}.  Jules Gavarret, in his 1840 book on medical statistics, cited Poisson's authority in deciding that 2 moduli gives ``a probability after which any therapeutic fact can and should be accepted without dispute''; see \S\ref{subsec:Gavarret}.

Laplace's theory was brought into English in the 1830s by Augustus De Morgan and Thomas Galloway, both of whom published books containing proofs of Laplace's theorem.  De Morgan's book appeared in 1837 \cite{De-Morgan:1837} and Galloway's in 1839 \cite{Galloway:1839}.  Galloway's may have been more influential among mathematicians, because whereas De Morgan followed Laplace directly, Galloway followed Poisson's simplified and clearer proof.  Both used the modulus.  So far as I know, De Morgan did not single out a particular number of moduli, but Galloway did mention 3 moduli; see \S\ref{subsec:Galloway}.

Cournot, in his 1843 book \cite{Cournot:1843}, followed Fourier in treating the probability of 1 in 20,000, corresponding to $\pm 2.87$ moduli, as practical certainty.  He did not round 2.87 to 3.  In \S35 of the book, he recommended that the limit 2.87 be held in mind not only because it corresponds to a value of P equal to 1 in 20,000 but also because it comes very close to 6 probable errors. In \S69, he called 2.87 moduli the ``limite extr\^eme de l'\'ecart''---the extreme limit of deviation.%
\footnote{This terminology may have already been established in the 1820s.  In his 1826 report, Fourier mentions that the difference between an observed and a true mean is called the ``erreur ou écart'' \cite[p.~3]{Fourier:1826}.}
Unlike Poisson and most of the earlier authors, Cournot explicitly rejected Laplace's Bayesian theory, accepting only a Bernoullian interpretation of the bounds given by Laplace's mathematics.  This makes his 1843 book very close to mathematical statistics as it was taught in the middle of the 20th century; the basic concepts are all there.  Cournot's ``limite de l'\'ecart'' was used in 20th-century French teaching of statistics until after World War II.

Although they used the modulus, Laplace, Fourier, Poisson, Gavarret, Galloway and Cournot did not use the name \emph{modulus} for it.  This usage apparently first appeared in English, in George Biddell Airy's 1861 book on the theory of errors \cite{Airy:1861}; see \S\ref{subsec:Airy}.  

Wilhelm Lexis, a prominent economist and statistician, kept the Laplacean tradition alive in Germany.  Although he taught and wrote in German, Lexis had studied for ten years in France, and the Laplacean aspiration to find practical certainty by multiplying observations was natural to his area of study. In his introduction to population statistics, \emph{Einleitung in die Theorie der Bev\"olkerungsstatistik} \cite{Lexis:1875}, published in 1875, Lexis repeatedly used 3 moduli as his level for practical certainty; see \S\ref{subsec:Lexis}.

\subsection{Tables of the normal distribution}

As mentioned in \S\ref{subsec:LG}, Bessel had calculated his small table of normal probabilities using Kramp's table of the incomplete integral of $e^{-t^2}$.  By the 1830s, it became common for books on probability to provide much more extensive tables of normal probabilities.  The first such table, also calculated from Kramp's, was given by the German astronomer Johann Franz Encke, a follower of Gauss, in an 1832 article on least squares \cite{Encke:1832}.  Encke tabulated the values of
\begin{equation}\label{eq:encke}
                \frac{2}{\sqrt{\pi}}\int_0^\tau e^{-t^2}dt
\end{equation}
to seven decimal places for values of $\tau$ (the number of moduli) from $0.00$ to $2.00$, in intervals of $0.01$.  He also gave a similar table in terms of the probable error.  
Encke's article was translated into English and printed with its tables in \emph{Taylor's Scientific Memoirs}, 1841, Vol.\ II, pp.\ 317--669 \cite[p.~180]{Merriman:1877}.  In 1838 \cite{De-Morgan:1838}, De Morgan extended Encke's tables from 2 to 3 moduli.  Galloway, in his 1939 book, also gave a table going up to 3 moduli.

In his 1843 book \cite{Cournot:1843}, Cournot gave a table of \eqref{eq:encke} for values of $\tau$ from $0.00$ to $3.00$, with a variable number of decimal places.  
He also gave the value for $\tau=3.00$ to $10$ places, for $\tau=4.00$ to $13$ places, and for $\tau=5.00$ to $17$ places.

\subsection{Outlying observations}

In the late 19th-century, United States scholars looked to Germany for intellectual leadership, and many of those interested in probability enlisted in the Gaussian school of least squares.%
\footnote{See, for example, the quotation by Mansfield Merriman in \S\ref{subsec:Merriman}.}
On one topic to which they made original contributions, however, the Americans brought Gaussian least squares into closer contact with Laplacean practical certainty.  This topic was the rejection of discordant or outlying observations.  

The best known suggestions were made by Benjamin Peirce in 1852 and William Chauvenet in 1863 \cite{Stigler:1973}.  Chauvenet suggested that the observation that deviates most from the average of $m$ observations be rejected if the probability of any particular observation deviating as much is less than $1/2m$.  Peirce's and Chauvenet's proposals were strongly criticized by Airy and other European experts on least squares.  In 1917, David Brunt reported that their criteria for rejection had not been widely used.  (Brunt's assessment is quoted more fully in \S\ref{subsec:Brunt}.)  Consideration of the problem of outliers forces us, however, to think about about how the probability of extreme or otherwise unlikely values increases as we look for them, and this was one of the ways the statistical literature in English began to acknowledge the problem of p-hacking.

\subsection{Edgeworthian \emph{significance}}\label{subsec:Edge}

A British economist and statistician, Edgeworth was by all accounts the first to use the English word \emph{significant} in connection with statistical testing.  He did this in a paper he read at the Jubilee meeting of the Statistical Society of London in 1885 \cite{Edgeworth:1885}.  The substance of the paper, as Edgeworth conceded in the discussion after the reading, was largely borrowed from Lexis.%
\footnote{Had Edgeworth studied mathematics at university, as Karl Pearson did, he might have learned the Laplacean theory from Galloway's book.  But as he had been trained as a classicist and was self-taught in mathematics, it would have been natural for him to seek the latest wisdom from a German authority.} 
Edgeworth's originality lie in translation; where Lexis discerned a real difference being \emph{praktisch gewiss} (practically certain), Edgeworth discerned an apparent difference \emph{signifying} a real difference; see~\S\ref{subsec:Edgeworth}.

An observed difference is significant in the sense of signifying if and only if two conditions are satisfied:  there is a real difference, and the observed difference is large enough to suggest it.  Either both conditions are satisfied, or not.  We may not be sure.  So when using the word in this sense, we may say that a difference is ``perhaps significant'', ``likely significant'', ``probably significant'', ``definitely significant'', or ``certainly significant''.  We may also say flatly that a difference is ``not significant''; if it is less than a probable error from zero, then it does not signify a real difference even if there is one.  We will not say that a difference is ``barely significant'' or ``just significant'', because if it does not definitely signify, then it may not signify at all.  Nor will we use phrases like ``highly significant'', ``very significant'', and ``more significant''.  It is the likelihood of signifying, not signifying itself, that is a matter of degree.

Edgeworth was anything but consistent, and as his paper rambled on, we do find one occasion where he writes ``highly significant''.  But for the most part he used ``significant'' to mean ``signifying'' something causal and not accidental dominated.  This usage seems odd to this speaker of American English in 2020.  Perhaps it was natural for Edgeworth's social class in his time and place, or perhaps it was merely one of his quirky turns of phrase.%
\footnote{Steve Stigler discusses Edgeworth's odd style on pp.~95--97 of \cite{Stigler:1999}.}
But Karl Pearson and his disciples understood it and adopted it.  They used it for thirty years or more.  As documented in \S\ref{subsec:Pearson}, it persisted into the 1920s in Pearson's \emph{Biometrika}.  The United States biometrician Raymond Pearl, a student of Pearson's, explained it very clearly in a book he published in 1923; see \S\ref{subsec:Pearl}. 

Pearson diverged from Edgeworth in an important respect; he measured the deviation of an estimate from the quantity estimated using the probable error rather than the modulus.  The most common formulation among the biometricians, reported by Pearl and followed by \emph{Biometrika}, was that a deviation of 3 probable errors (about two standard deviations) was likely significant and a deviation of 6 probable errors (about 4 standard deviations) was definitely significant.

\subsection{Enter the American psychologists}

Edgeworthian significance disappeared in the 1920s.  The word remained, but the meaning shifted.  This seems to have been a gradual process, unnoticed by many.  It seems that many statisticians outside Pearson's international circle of biometricians picked up the word \emph{significance} without grasping its Edgeworthian interpretation, which must have been as unexpected for many ears then as it is for mine now.

One field where we can see this happening is psychology.  As Steve Stigler has noted \cite{Stigler:1992}, psychologists had begun using mathematical statistics in the 1860s and had developed their own methods long before Pearson created biometry.  In the late 1910s and late 1920s, we see young United States psychologists using \emph{significant} and \emph{significance} in relatively vague and certainly non-Edgeworthian ways.  In 1922, for example, we find Morris Viteles, who later became very prominent in industrial and organizational psychology, writing that test results were ``greatly significant'' and ``highly significant''.  He may also have been the first to use \emph{level of significance} in connection with statistical testing; see \S\ref{subsec:Viteles}.

The first use of the non-Edgeworthian term \emph{statistical significance} in its modern sense that I have found is in a 1916 article by another young and eventually very prominent U.S.\ psychologist, Edwin Boring \cite[p.~315]{Boring:1916}.  Boring understood the vocabulary of the British biometricians reasonably well, but he soon concluded that the assumptions underlying the Laplacean method (e.g., independence of observations and a common meaning for a parameter in different individuals or groups) were usually not satisfied in his work.  His most often cited criticism of the method was a 1919 article entitled ``Mathematical vs.\ scientific significance'' \cite{Boring:1919}.  He carried on a years-long debate on the use of statistics in psychology with Truman Kelley, at the time one of psychology's most prominent experts on statistical methods \cite{Stout:1989}.

In his 1923 textbook \emph{Statistical Method} \cite{Kelley:1923} Kelley wrote, ``If these two relationships do not exactly hold, the significance of the discrepancy can be determined by formulas giving probable errors\ldots'' (p.~99).  This vague assertion is not quite Edgeworthian, for the comparison of an estimate with its probable errors often leaves us uncertain whether it signifies in the Edgeworthian sense.  On page~102, at the beginning of a lengthy passage quoted in \S\ref{subsec:Kelley}, Kelley made a similar statement:  ``The significance of any measure is to be judged by comparison with its probable error.'' This passage is also of interest because it shows how Kelley was shifting his readers from the probable error to the standard deviation, and because it shows how to perform a one-sided test.

Shortly before Kelley competed his book, he had spent a sabbatical year in London with Pearson.%
\footnote{Personal communication from Lawrence Hubert, who has examined the Kelley archive at Harvard.  See also \cite{Bellhouse:2009}.}
Perhaps he also met Fisher at that time.  When he sent Fisher a copy of the text, Fisher responded that it was ``quite the most useful and comprehensive book of the kind yet written'' (1924, January 12) \cite[p.~560]{Stout:1989}.

\subsection{Fisher's \emph{Statistical Methods for Research Workers}}\label{subsec:Fisher1}

In 1925, Fisher published his own statistics manual, his celebrated \emph{Statistical Methods for Research Workers} \cite{Fisher:1925}.  The words \emph{significant} and \emph{significance} are prominent in this book, but their Edgeworthian meaning has slipped away in favor of a meaning that allows degrees of significance.   Probable error has given way completely to standard deviation.

The main purpose of the book was to provide tables for the many distributions that Fisher had studied, including Student's $t$ and the distribution of the correlation coefficient, and to teach research workers how to use these tables.  Because these distributions were not normal and sometimes not symmetric, ``significance'' could not be defined in terms of standard deviations.  Fisher instead defined it directly in terms of tail probabilities.  Two standard deviations was replaced by 5\% \cite{Stigler:2008}.

None of these features was unprecedented.  Edgeworth's \emph{significance} was already fading in some quarters.  Yule had emphasized standard errors in his popular 1911 textbook (see \S\ref{subsec:Yule}).  We saw Fourier mentioning odds of 19 to 1 in 1826, and in 1919 David Brunt used these odds in another context where practical certainty could not be measured in terms of probable errors (see \S\ref{subsec:Brunt}).

Fisher's tone does not suggest that he had deliberated about rejecting the Edgeworthian meaning of \emph{significant}.  He was never part of Pearson's circle, and by 1925 he was certainly not looking to Pearson's work for guidance.  It seems likely that he drew his vocabulary less from \emph{Biometrika} than from the U.S. psychologists or others distant from Pearson.  Perhaps this included ``research workers'' at the agricultural experiment station at Rothamsted, where he had already been working for five years.

\subsection{Seeing p-hacking in Britain}\label{subsec:acknowledge}

As soon as we have significance testing, we have p-hacking.  We can find Laplace himself varying his method of testing in search of a small p-value.  Poisson searched across categories to find apparently significant ratios of male to female births.  In the early volumes of \emph{Biometrika}, we can find authors giving lists of estimated differences and their probable errors and then calling attention to one or two cases where the ratio of the estimated difference to its probable error appears to signify.  But what makes this p-hacking visible to statisticians?  Do we need a mathematical philosopher like Cournot to see it?

If you begin, as many 19th- and early 20th-century British mathematicians did, by assuming that probability is a relation between proposition and evidence, then you may find it paradoxical that the search producing the evidence should matter, and this may make it difficult to see p-hacking.  But the historical record reveals at least two cases where the search was too obvious for some British mathematicians to ignore.  The first was the problem of rejecting discordant observations (outliers); the second was the search for cycles in time series using harmonic analysis.

\paragraph{Discordant observations.}

The erudite Edgeworth knew Cournot's work.  When he addressed the problem of discordant observations (outliers) in 1887, Edgeworth drew on Cournot's insights to understand the need to take account of the number of observations in deciding whether extreme observations should be considered discordant.  Cournot's view was paradoxical, Edgeworth thought, but right (\cite[pp.~369--370]{Edgeworth:1887} quoted in \S\ref{subsec:Edgeworth}).  The next year, John Venn, in his \emph{Logic of Chance}, acknowledged Cournot's and Edgeworth's point in the course of a discussion of fallacies in probability judgement \cite[3rd edition, pp.~338--339]{Venn:1866}. 

I have not found found examples before the 1920s in which Pearson or his disciples cited Cournot's insights concerning p-hacking or Edgeworth's related insights concerning discordant observations.  Perhaps they were unaware of these insights and unaware in general of the hazards of p-hacking.  But Pearson certainly knew Edgeworth's work, and silence in print about a problem that one cannot solve is not quite evidence of lack of awareness.  The alternative hypothesis, that Pearson's circle did talk about p-values being invalidated by search or selection, is supported by a passing reference to the problem by Pearson's son, Egon S.\ Pearson, in 1925:  Egon called it``the old difficulty''.    A decade later, when Egon and Chidambara Chandrasekaran noted the illegitimacy of choosing a test after seeing the data, the context was precisely Edgeworth's:  the rejection of outliers.  (See \S\ref{subsec:Pearson-ES} for references and fuller quotations.)

\paragraph{Searching for cycles.}

In the case of cycles, it was the meteorologists who instructed the biometricians. 

The notion that cycles in time series might be discovered and used for prediction was very popular at the end of the 19th and beginning of the 20th centuries, when William Stanley Jevons and other respected scholars even conjectured a causal chain from sunspots to business cycles:  cycles in sunspots might cause cycles in the weather, hence cycles in agricultural production and other economic activity \cite{Morgan:1990,Klein:1997,Friedman:2014}.  A statistical test for whether an apparent cycle in a time series is real was suggested in 1898 by the British physicist Arthur Schuster, in the article in which he introduced the name \emph{periodogram} for a graph showing the estimated intensity of different frequencies in the series' Fourier transform \cite{Schuster:1898}.  Schuster eventually explained his test in a very simple way:  the probability that a particular estimated intensity will be $h$ or more times as large as its expected value is $e^{-h}$.  (Being the sum of the squares of two normally distributed variables, it will have an exponential distribution; see \cite{Schuster:1906} and \S\ref{subsec:Schuster}.) 

Looking for cycles in a time series is a way of searching through the data for something remarkable.  One of the first, perhaps the very first, to point out how misleading this particular type of search can be was Gilbert T.\ Walker, a physicist working as a meteorologist in India; see \S\ref{subsec:Walker}.  Walker's first critique was published in India, in 1914 \cite{Walker:1914}.  The same point was made in 1919 by the physicist F.\ J.\ W.\ Whipple in the discussion of a paper on cycles in the Greenwich temperature records, read to Royal Meteorological Society by David Brunt  \cite{Brunt:1919}; see \S\S\ref{subsec:Brunt} and \ref{subsec:Whipple}.

The British biometricians may have overlooked Walker's 1914 critique and Whipple's 1919 comments.  But there is no doubt that the p-hacking issue raised by Schuster's test came to their attention in 1922, after the prominent civil servant and scholar William Beveridge read a paper to the Royal Statistical Society on cycles in rainfall and wheat prices.  Beveridge read his paper, in which he more or less used Schuster's test, on April 25 of that year \cite{Beveridge:1922}.  Yule was one of the discussants.  None of the Society members who commented on the paper were fully convinced by Beveridge's conclusions, but he stirred their interest.  In its issues for August 19 and August 26, (vol.~110, pp.~265, 289), \emph{Nature} reported that the fall meeting of the British Association at Hull would include a special session on ``Weather Cycles in Relation to Agriculture and Industrial Fluctuations", to be held on September 7.  The session was to be sponsored jointly by three sections of the association, Economic Sciences and Statistics, Mathematics and Physics, and Agriculture, and it was to feature discussion by Beveridge, Fisher, and Yule.   In its December 30 issue (vol.~110, pp.~889--890), \emph{Nature} summarized the discussion.  Beveridge made his case for relating periodicities in the weather to those of wheat prices.  Yule mildly defended Beveridge against those who found his case entirely implausible.  Fisher questioned how strongly periodicities in the weather could be related to periodicities in production, reporting that the total rainfall at Rothamsted accounted for a relatively small amount of the variation in production. 

Walker re-entered the story before December 30.  In its October 14 issue (vol.~110, pp.~511--512), \emph{Nature} published a letter to the editor from Walker, along with a response by Beveridge \cite{Walker:1922}; see \S\S\ref{subsec:Walker} and \ref{subsec:Beveridge}.  Walker made the point that he had made already in 1914:  if a statistician is going to test the largest estimated intensity from a periodogram that shows estimated intensities for $k$ different frequencies, the probability of this greatest estimated intensity being $h$ or more times as large as the expected value for any given intensity is $e^{-h/k}$ rather than $e^{-h}$.  With this adjustment, Walker did not find Beveridge's cycles to be convincing.  Beveridge conceded the conceptual point but adjusted the assumptions in Walker's analysis so that his own conclusions emerged intact.

What was Fisher to say about this?  The whole periodogram story being a mess, he probably did not have much to contribute, and nothing would have been gained by entering a controversy between two such powerful individuals.  In due time, however, Fisher did address the problem of selecting the most significant test.  In 1925, when he finally published his thoughts on the Rothamsted data on rainfall and crop production \cite[p.~94--95]{Fisher:1925b}, he derived the adjustment required when one variable is chosen from several for a regression.  In 1929, he showed how Schuster's and Walker's criteria for testing intensities in a periodogram can be adjusted to account for the fact that the variance must be calculated from the data \cite{Fisher:1929}.   
 
There are, however, no cautions about p-hacking in Fisher's first edition of \emph{Statistical Methods}.  Why did he omit the topic?  The obvious answer is that the topic is impossibly difficult for a book that offers research workers with limited mathematical training recipes for statistical testing.  Perhaps too difficult for anyone.  As Beveridge's response to Walker's 1922 letter suggests, adjusting p-values for selection is often topic for debate, not for recipes.

In the preface to the sixth edition of \emph{Statistical Methods}, published in 1936 (p.~xii), we finally see a recipe of sorts for dealing with selection, gingerly offered:  perhaps a very high level of significance, such as $0.1$ per cent, should be used; see \S\ref{subsec:Fisher}.  In this same preface, Fisher refutes critics who had asked why he did not provide mathematical derivations for his recipes in the book. The book, he explains, is for research workers, not people doing mathematical theory.%
\footnote{I am indebted to John Aldrich for calling my attention to Fisher's 1925 article and this preface.}

\subsection{Who invented the name \emph{p-value}?}\label{subsec:pvalue}

No one.  The term simply evolved from the use of the letter P to denote the probability that an estimated quantity or difference will fall inside or outside given limits.

We already see this use of P, in Roman upper case, in Fourier, Poisson, Gavarret, and Cournot.  Beginning at least in his 1900 article on $\chi^2$ \cite{Pearson:1900}, Karl Pearson similarly wrote P for the probability of a result more extreme than the observed value of a test statistic and referred to it as ``the value of P''.  Yule and Fisher followed Pearson's example throughout their careers \cite{Yule:1911,Fisher:1925a,Fisher:1948}.  

By the 1920s, some authors outside Britain had occasionally and casually turned \emph{value of P} into \emph{P value}.  The earliest examples I have seen are in articles by the North American geneticist John W. MacArthur in 1926 \cite[pp.~397, 400]{MacArthur:1926}, the United States biostatistician Persis Putnam in 1927 \cite[pp.~672--673]{Putnam:1927}, and the Chinese statistician C.\ P.\ Sun in 1928 \cite[p.~67]{Sun:1928}.%
\footnote{I am indebted to Sander Greenland for calling my attention to these articles.}  
Later authors who used \emph{P value} or \emph{P-value} casually include John Wishart \cite[p.~304]{Wishart:1938} and his associate  H.\ O.\ Hirschfeld in 1937 \cite[p.~68]{Hirschfeld:1937}%
\footnote{Hirschfeld later anglicized his name to H.\ O.\ Hartley.}  
and W.\ Edwards Deming in 1943 \cite[p.~30]{Deming:1943}.  None of these authors gave any indication that they thought they were coining a novel usage.  I have not yet seen any use of \emph{P value} in the social sciences before 1970; we do not see it in any of the articles in \cite{Morrison/Henkel:1970}.  

Today the use of \emph{P-value} is widespread, but there is no consensus on the font for the letter P.  We see lower and upper case, italics and roman, text and mathematical font, with and without the hyphen.

\section{Conclusion}\label{sec:conclusion}

In 1949 \cite[p.~90]{Wishart:1949}, the accomplished British statistician John Wishart wrote, 
\begin{quote}
If one were asked to say what has been the distinctively British contribution to the theories of probability and mathematical statistics during the present century, the answer, I fancy, would be found, not so much in the formulation of a satisfactory theory of probability, including the nature of inference, as in the fashioning of significance test tools to guide the practical experimenter.
\end{quote}
The history reviewed in this paper confirms Wishart's judgement.  The notions of testing and estimation used in mathematical statistics even today were in place already in the 19th century.  

There is also a parallel with respect to the misuse and abuse of these basic concepts.  The inappropriate models and inferences that led to the collapse of the Laplacean tradition in France in the second half of the 19th century are rampant today, inspiring loss of confidence and hand-wringing. We see a blizzard of proposals to correct these problems.  Some propose to shift the level required for calling attention to a p-value (replace Fisher's 0.05 with Poisson's and Gavarret's 0.005); some propose to change the words we use (eliminate \emph{p-value} or \emph{statistical significance}); others propose various more or less complicated ways of complementing the statement of a p-value.  

What lessons should we draw from the durability of these basic ideas and of their abuses?  Everyone will answer this question for themselves, but the durability does suggest that superficial changes in terminology will not be helpful.  The quick emergence of significance testing and p-values from Laplace's 19th-century confidence intervals further suggests that urging people to return to confidence intervals and stay there may be equally futile.

The failure of 20th-century embellishments to forestall misunderstanding and abuses also suggests that we might well simplify our elementary teaching by emphasizing Fourier's and Cournot's Bernoullian calculation of limits on error based on the central limit theorem, treating p-values for other tests as an intuitive extension of this basic example.  This simplification would allow space in the curriculum to emphasize the inevitability of p-hacking and the necessity of treating some calculations as merely exploratory analysis.  It would also allow some space to teach about testing by betting.

\subsection{Can we return to practical certainty?}\label{subsec:returnpc}

In March 2019, hundreds of statisticians lent their support to a commentary in \emph{Nature} entitled ``Retire statistical significance'' \cite{Amrhein/etal:2019}.  What should replace it?  How is  a scientist or journalist inexpert in statistical mathematics to interpret a p-value?

Here is a fanciful thought experiment.  Suppose we all (all the teachers and textbook writers in statistics) reach a consensus to return to Edgeworthian significance.  And suppose we signal this change by replacing \emph{significant} with \emph{signifying}.  When a test is fully planned in advance,
\begin{itemize}
\item
a p-value of 0.05 (about three probable errors or two standard deviations) is likely to signify; 
\item
a p-value of 0.005 (about four probable errors or two moduli) is practically certain to signify;
\item
a p-value of 0.00006 (about six probable errors or four standard deviations) definitely signifies.  
\end{itemize}
What problems would this pose?  The obvious problem is the ``likely'' in ``likely to signify''.  Shall we give it a Bayesian interpretation with a uniform prior, as Edgeworth did?  A Bernoullian interpretation as Cournot and Jerzy Neyman would?  Here the fancied consensus splinters.

In my view, we do not particularly need Edgeworth's \emph{significant} or \emph{signify} in the teaching of elementary statistics, but there is much to be said for returning to the simple language of Bernoullian practical certainty as it was used by Cournot in 1843.  Leaving aside any appeal to an axiomatic theory of probability, we could teach students how to find reasonable limits on error, and then use these limits to obtain confidence intervals, just as Cournot, Neyman, and countless statisticians between them did.  Then, following Gavarret, we could translate this into a two-sided test for a difference between two treatments: one treatment is practically certain to be better than the other when its observed advantage is greater than the limit of possible error; otherwise we are not authorized to pronounce in favor of either treatment; see \ref{subsec:Gavarret}.

When Jacob Bernoulli first undertook to base judgements of \emph{probability} in civic, moral, and business matters on calculations like those in games of chance, he did not envision counting chances for and against every civil, moral, and business question.  He certainly did not suppose that every proposition would have probabilities for and against adding to one.  On the contrary, he thought chances could be counted only for some propositions, and he hoped that arguments that compared and combined these counts could then cast light on other propositions.  This might not produce probabilities that follow Kolmogorov's axioms, but it might sometimes lead to practical certainty \cite{Shafer:1978}.   Gavarret's explanation of how to use a ``limit of possible errors'' stands squarely in this Bernoullian tradition, which may allow us to conclude that a proposition is practically certain without assigning it a numerical probability.

Critics of the current pedagogy of elementary mathematical statistics have complained that it has become a confusing mixture of competing theories and ideas, mixing Fisherian, Bayesian, and decision-theoretic viewpoints \cite{Denis:2004,Gigerenzer:2018,Hurlbert/etal:2019}.  To a large extent, these competing theories are different ways of forcing the judgements of practical certainty taught by Laplace, Fourier, and their 19th-century successors into the 20th-century's measure-theoretic framework for mathematical probability.  Must we do this?

\subsection{Can we return to exploratory data analysis?}

Beginning in the late 1940s, when the center of research in mathematical statistics had shifted to the United States, there was a great deal of work on formal methods for \emph{multiple comparisons}---a name appropriate when a plan is made in advance to test many different hypotheses and significance levels are adjusted accordingly \cite{Shaffer:1995,Mayo:2018}.  John Tukey, one of the leaders in this work \cite{Benjamini/Braun:2002}, was acutely aware that it was not always relevant to practice.  In practice, scientists often cannot plan in advance what they will think of doing later.  One way of dealing with this problem is to drop the ambition of arriving at practical certainties in a particular study, treating the study as purely exploratory.  With this in mind, Tukey coined the term \emph{exploratory data analysis}, to be distinguished from \emph{confirmatory data analysis}.  Promising results from an exploratory study were not to be taken seriously as science until they were confirmed by later studies.

The sociological realities of academia have not been kind to exploratory data analysis.  It has not proven to be a path to publication and prestige, and it is now often thought of as merely a collection of techniques for the visual and tabular presentation of data.  Some contributors to recent discussions of significance testing in \emph{The American Statistician} have suggested putting new emphasis on the exploratory/confirmatory contrast \cite{Wasserstein/etal:2019}.  If we were to simplify the discussion of testing by returning to Gavarret's simple framework, there might be room for this in elementary statistics instruction.

\subsection{Can we return to betting?}

At its core, probability theory is a calculus of betting, but in the three centuries since Bernoulli, mathematical statistics has been increasingly formalized and taught in ways that hide this fact.  My own proposal for fundamental change in statistical testing is to bring betting back to the center of the picture.

To do this, we must shift from the Bayesian statement that
\begin{quote}
assuming model B, the observations make hypothesis A unlikely
\end{quote}
and the Bernoullian statement that
\begin{quote}
assuming B, the observations are unlikely if A is true
\end{quote}
to a statement about a bet:
\begin{quote}
a bet that was fair according to B has paid off handsomely (multiplied the money it risked by a large factor) if A is true.
\end{quote}
When we do this, we may say that we are testing the hypothesis A by betting against it.  This has two obvious advantages.  First, it makes very salient the need to state the bet in advance; no one is impressed when you identify after the fact a bet that would have been successful.  Second, it makes very salient the remaining uncertainty; no one forgets that success in betting, no matter how striking, may have been accidental.

As I argue in \cite{Shafer:Toappear}, testing by betting can also help us give more honest accounts of opportunistic searches for significance.  A honest Bayesian or Bernoullian account of such a search requires the specification of a complete strategy for the search.  What would you have done if the first test had come out differently, etc.?  If the search is opportunistic, such a plan can only be conjectured after the fact and can hardly ever be convincing.  When we test by trying to multiply the money we risk, no grand strategy is required; we can change direction opportunistically so long as each successive bet is honest (made before we look) and risks only the net capital resulting from the preceding bet.

\section{Acknowledgements}

This paper would not have been possible without insights freely shared by John Aldrich and Bernard Bru.  I have also benefited from conversations with many other colleagues, especially Michel Armatte, Hans Fischer, Steve Goodman, Prakash Gorroochurn, Sander Greenland, Deborah Mayo, Aris Spanos, Steve Stigler, Volodya Vovk, and Sandy Zabell.

\begin{appendices}

\section{What they said}\label{app:FC}

In this section, I provide more detail about how a number of 19th and early 20th century authors wrote about Laplace's theorem, error limits, and practical certainty.  Most of these authors worked in the Laplacean tradition.  I consider them in order of their birth.

I have chosen these authors, somewhat arbitrarily, to illustrate how the Laplacean tradition evolved as it was transmitted into English.  A more comprehensive review would include authors working in a wider variety of applications and in other European countries, including Italy, Belgium, and Denmark.

In translations, I have sometimes shifted the authors' notation to current practice in American English, italicizing symbols, indicating limits of integration with subscript and superscript, writing $0.9953$ instead of 0,9953 or 0$\cdot$9953, writing $h^2$ instead of $hh$, etc.

\subsection{Joseph Fourier, 1768--1830}\label{subsec:Fourier}

Joseph Fourier is most renowned for his mathematical analysis of the diffusion of heat, but he was also a revolutionary and a politician, an impassioned participant in the French revolution and an administrator under Napoleon.  After Napoleon's final defeat and the return of a royalist regime in 1815, Fourier was briefly left with neither a position nor a pension, but the royalist Chabrol de Volvic, who had been his student, rescued him from impoverishment with an appointment to the census bureau of the Paris region \cite{Ozouf-Marignier:1997}.  The appointment left him time for his mathematical research, but he faithfully attended to his duties at the census, issuing masterful reports in 1821, 1823, 1826, and 1829.  Fourier's name was not included in the reports, but there is no doubt that he edited them and wrote the mathematical memoirs that appear at the beginning of the 1826 and 1829 ones \cite[p.~198]{Bru/etal:1997}.

Given independent observations $y_1,\ldots,y_m$ and their average $\overline{y}$, Fourier estimated what I am calling the modulus by
\[
          g := \sqrt{\frac{2}{m}\left(\frac{\sum_{i=1}^m y_i^2}{m} - \overline{y}^2\right)}
\]
This is consistent with modern practice; the modulus is $\sqrt{2}$ times $\overline{y}$'s standard deviation, and $g$ is $\sqrt{2}$ times the maximum likelihood estimate of this standard deviation.

The passage from Fourier's 1826 memoir translated here includes a table of significance levels, which may look more familiar when we add a third column translating units of $g$ into units of $\overline{y}$'s standard error $g/\sqrt{2}$:
\begin{center}
\def\arraystretch{1.2}
\begin{tabular}{C{3cm}C{1cm}C{3cm}}
    units of $g$ &  P &  units of $g/\sqrt{2}$\\
     0.47708   & $\frac12$ & 0.67\\
     1.38591   & $\frac{1}{20}$ & 1.96\\
     1.98495   & $\frac{1}{200}$ & 2.81\\
     2.46130   & $\frac{1}{2000}$ & 3.48\\
     2.86783   & $\frac{1}{20000}$ & 4.06
\end{tabular}
\end{center}
Fourier wrote that it is ``a 19 out of 20 bet'' that the error will not exceed $1.38591 g$.  This is the familiar 95\% confidence interval obtained using 1.96 standard errors.  He also concludes that the error certainly will not exceed $3g$. 

The following passage constitutes \S XI of the 1826 memoir \cite[pp.~xxi--xxii]{Fourier:1826}.

\subsubsection*{Fourier in English}

To complete this discussion, we must find the probability that H, the quantity sought, is between proposed limits $\text{A} + \text{D}$ and $\text{A} - \text{D}$.  Here A is the average result we have found, H is the fixed value that an infinite number of observations would give, and D is a proposed quantity that we add to or subtract from the value A.  The following table gives the probability P of a positive or negative error greater than D; this quantity D is the product of $g$ and a proposed factor $\partial$.

\begin{center}
\def\arraystretch{1.2}
\begin{tabular}{cc}
     $\partial$ &  P\\
     0.47708   & $\frac12$ \\
     1.38591   & $\frac{1}{20}$ \\
     1.98495   & $\frac{1}{200}$ \\
     2.46130   & $\frac{1}{2000}$ \\
     2.86783   & $\frac{1}{20000}$ 
\end{tabular}
\end{center}

Each number in the P column tells the probability that the exact value H, the object of the research, is between $\text{A} + g\partial$ and $\text{A} - g\partial$.  Here A is the average result of a large number $m$ of particular values $a, b, c, d,\ldots, n$; $\partial$ is a given factor; $g$ is the square root of the quotient found by dividing by $m$ twice the difference between the average of the squares $a^2,  b^2, c^2, d^2,\ldots,n^2$ and the square A$^2$ of the average result. We see from the table that the probability of an error greater than the product of $g$ and 0.47708, i.e. greater than about half of $g$, is $\frac12$.  It is a  50--50 or 1 out of 2 bet that the error committed will not exceed the product of $g$ and 0.47708, and we can bet just as much that the error will exceed this product.

The probability of an error greater than the product of $g$ and 1.38591 is much less; it is only $\frac{1}{20}$. It is a 19 out of 20 bet that the error of the average result will not exceed this second product.

The probability of an even greater error becomes extremely small as the factor $\partial$ increases.  It is only $\frac{1}{200}$ when $\partial$ approaches $2$.  The probability then falls below $\frac{1}{2000}$.  Finally one can bet much more than twenty thousand to one that the error of the average result will be less than triple the value found for $g$.  So in the example cited in Article VI, where the average result was $6$, we can consider it certain that the value $6$ is not wrong by a quantity three times the fraction $0.082$ that the rule gave for the value of $g$.

The quantity sought, $H$, is therefore between $6-0.246$ and $6+0.246$.

\subsubsection*{The French original}

Pour compléter cette discussion , il faut déterminer quelle probabilité il y a que la quantité cherchée H est comprise entre des limites proposées $\text{A} + \text{D}$ et $\text{A} - \text{D}$.  A est le résultat moyen que l'on a trouvé, H est la valeur fixe que donnerait un nombre infini d'observations , et D est une quantité proposée que l'on ajoute à la valeur A ou que l'on en retranche. La table suivante fait connaître la probabilité P d'une erreur positive ou négative plus grande que D; et cette quantité D est le produit de $g$ par un facteur proposé $\partial$.

\begin{center}
\def\arraystretch{1.2}
\begin{tabular}{cc}
     $\partial$ &  P\\
     0,47708   & $\frac12$ \\
     1,38591   & $\frac{1}{20}$ \\
     1,98495   & $\frac{1}{200}$ \\
     2,46130   & $\frac{1}{2000}$ \\
     2,86783   & $\frac{1}{20000}$ 
\end{tabular}
\end{center}

Chacun des nombres de la colonne P fait connaître quelle probabilité il y a que la valeur exacte H, qui est l'objet de la recherche , est comprise entre les limites $\text{A} + g\partial$ et $\text{A} - g\partial$.  A est le résultat moyen d'un grand nombre $m$ de valeurs particulières $a, b, c, d,\ldots, n$; $\partial$ est un facteur donné; $g$ est la racine carrée du quotient que l'on trouve en divisant par $m$ le double de la différence de la valeur moyenne des carrés $a^2,  b^2, c^2, d^2,\ldots,n^2$ au carré A$^2$ du résultat moyen.  On voit par cette table que la probabilité d'une erreur plus grande que le produit de $g$ et 0,47708, c'est-à-dire, plus grande qu'environ la moitié de $g$, est $\frac12$.  II y a 1 contre 1 ou 1 sur 2 à parier que l'erreur commise ne surpassera pas le produit de $g$ par 0,47708, et il y a autant à parier que l'erreur surpassera ce produit.

La probabilité d'une erreur plus grande que le produit de $g$ par 1,38591 est beaucoup plus petite que la précédente; elle n'est que $\frac{1}{20}$. II y a 19 sur 20 à parier que l'erreur du résultat moyen ne surpassera pas ce second produit.

La probabilité d'une erreur plus grande que la précédente devient extrêmement petite, à mesure que le facteur $\partial$ augmente. Elle n'est plus que $\frac{1}{200}$ lorsque $\partial$ approche de $2$. La probabilité tombe ensuite en dessous de $\frac{1}{2000}$. Enfin, il y a beaucoup plus de vingt mille à parier contre 1 que l'erreur du résultat moyen sera au-dessous du triple de la valeur trouvée pour $g$. Ainsi , dans l'exemple cité art.\ VI, où l'on a 6 pour le résultat moyen, on peut regarder comme certain que cette valeur 6 n'est pas en défaut d'une quantité triple de la fraction 0,082 que la règle a donnée pour la valeur de $g$.

La grandeur cherchée H est donc comprise entre $6 - 0,246$ et $6+0,246$.

\subsection{Sim\'eon-Denis Poisson, 1781--1840}\label{subsec:Poisson}

Poisson advanced Laplace's theory substantially.  Beginning in the 1820s, he simplified the proof of Laplace's theorem, making it accessible to many more mathematicians \cite[\S17.3]{Hald:1998}.  In 1830, he gave straightforward instructions for calculating limits of practical certainty for the difference between two proportions \cite{Poisson:1830}.%
\footnote{In his second memoir on mathematical statistics, in 1829 \cite{Fourier:1829}, Fourier had explained how to calculate limits on a function of several estimated quantities, but he had not spelled out how his formulas specialize to the case where this function is simply the difference between two proportions.}  
Finally, in 1837, he pulled together his theoretical and applied results on probability in an impressive treatise, \emph{Recherches sur la probabilit\'e des jugements} \cite{Poisson:1837}. 

Like Fourier, Poisson discussed limits in terms of numbers of moduli.  When writing theory, he required 3, 4, or even 5 moduli for practical certainty \cite[\S\S80, 87, and 96]{Poisson:1837}.  But when analyzing data, he used less exigent limits.  In \S89, when dealing with Buffon's data, he gave limits and odds corresponding to 2 moduli.  In \S111, he reduced this to 1.92 moduli, corresponding to a bet at odds 150 to 1.  

An example of a theoretical discussion is found in \S87, where Poisson considered the problem of testing whether the unknown probability of an event $E$ has changed between the times two samples are taken.  There are $\mu$ observations in the first sample; E happens in $n$ of them, and its opposite $\text{F}= \text{E}^c$ happens in $m=\mu-n$ of them.  For the second sample, he uses analogous symbols $\mu'$, $n'$, and $m'$.  He gives formulas, under the assumption that the unknown probability has not changed, for the estimated modulus of the difference
$
             \frac{m'}{\mu'} - \frac{m}{\mu}
$
and for the probability that this difference will be within $u$ moduli of $0$.  Then he writes,
\begin{quote}
So if we had chosen a number like three or four for $u$, making the probability $\tilde{\omega}$ very close to certainty (n$^\text{o}$ 80), and yet observation gives values for $\frac{m'}{\mu'} - \frac{m}{\mu}$ or $\frac{n'}{\mu'} - \frac{n}{\mu}$ that are substantially outside these limits, we will have grounds to conclude, with very high probability, that the unknown probabilities of the events E and F have changed in the interval between the two series of trials, or even during the trials.
\end{quote}
\begin{quote}
Si donc on a pris pour $u$ un nombre tel que trois ou quatre, qui rende la probabilit'e $\tilde{\omega}$ tr\`es approchante de la certitude (n$^\text{o}$ 80), et si, n\'eamoins, l'observation donne pour $\frac{m'}{\mu'} - \frac{m}{\mu}$ ou $\frac{n'}{\mu'} - \frac{n}{\mu}$ des valeurs qui s'\'ecartent notablement de ces limites,on sera fond\'e \`a en conclure, avec une tr\`es grande probabilit\'e, que les chances inconnu des \'ev\'evements E et F ont chang\'e, dans l'intervalle des deux s\'eries d'\'epreuves, ou m\^eme pendant ces \'epreuves.
\end{quote}

The closest Poisson came to identifying $\pm2$ moduli with practical certainty may have been in \S135 of the book, where he considered the 42,300 criminal trials in France during the years 1825 through 1830.  The defendant was convicted in 25,777 of these trials.  So his estimate of the average probability of conviction, which he called R$_5$, was $(42300/25777)\approx0.6094$.  His estimate of its modulus was $0.00335$.  He states that if we use 2 moduli, 
\begin{quote}
\ldots we will also have
\[
    \text{P}=0.9953,
\]
for the probability, very close to certainty, that the unknown R$_5$ and the fraction $0.6094$ will not differ from each other by more than $0.0067$.
\end{quote}
\begin{quote}
\ldots on aura aussi
\[
    \text{P}=0.9953,
\]
pour la probabilité, très approchante de la certitude, que l'inconnue R$_5$ et la fraction 0,6094 ne diffèrent pas de 0,0067, i'une de l'autre.
\end{quote}

\subsection{Friedrich Wilhelm Bessel, 1784--1846}\label{subsec:Bessel}

Having determined the position of over 50,000 stars, Friedrich Wilhelm Bessel was renowned as an astronomer.  In the course of his work, he developed and popularized Gauss's theory of errors.  He believed that systematic errors are often more important than random errors, and his influence helped establish the emphasis on perfecting instruments and computational methods that pushed the Germans ahead of the French in astronomy and geodesy by the middle of the 19th century.

The passage translated here is \S10 (pp.~141--142) of Bessel's study of the orbit of Olber's comet \cite{Bessel:1816}.  Published in 1816, it includes the first known tabulation of significance levels for the normal distribution.  The table shows the odds that an error will fall within $\pm \alpha\times(\text{probable error})$ for values of $\alpha$ up to $4$.  The odds are more than $140$ to $1$ that it fall within $\pm 4 \times (\text{probable error})$.  (Four probable errors is about two moduli or $2.8$  standard deviations.)  But Bessel does not pause over whether this should be regarded as practical certainty.  The point of the table is not to show what is required for practical certainty but to show why different observations (or equations) must be weighted differently in order to arrive at the best estimates of unknowns.

Bessel is credited with inventing the notion of a probable error.  In the translated passage he recommends estimating the probable error from the observations, taking it to be the median of the observed absolute errors.  

\subsubsection*{Bessel in English}

Success in determining the final values of quantities from these equations of condition, and even more so the estimation of their likely uncertainty arising from errors in the observations, depends principally on the proper weighting of the equations of condition. It was, therefore, necessary to make a study of this question, the result of which I have already used with advantage for some years.

According to Gauss's least-squares theory, the probability of making an error $\Delta$ is
\[
        \phi(\Delta) = \frac{h}{\sqrt{\pi}} e^{-h^2\Delta^2}
\]
(\emph{Theoria mot.\ corp.\ coel.\ P.\ 212.}), where $h$ depends on the precision of the observations.  By means of this expression one can easily determine the probable error of a single observation from an actual set of observations, under the assumption that the errors that actually occur are free from all systematic influences, and are produced only by the imperfections of the instruments and senses. Indeed, the greater the number of observations, the closer we come to the arithmetic mean of all errors, taken together with the same sign, which we shall call $\epsilon$,
\[
        = 2\int_0^\infty \phi(\Delta)\Delta d\Delta =\frac{1}{h\sqrt{\pi}};
\]
and also to the square root of the arithmetic mean of the squares of the errors, which we will denote by $\epsilon$, from the equation
\[
         \epsilon'^2 = 2 \int_0^\infty \phi(\Delta) \Delta^2d\Delta = \frac{1}{2h^2}.
\]

The greater the number of actual observations, the more we are entitled to assume that these errors occur as the Gaussian theory requires, so that from the coincidence of the $\epsilon$ and $\epsilon'$ obtained  from a very large number of observations with the best possible corresponding values from the theory, we now obtain the probable error of an observation, which we will denote $\epsilon''$. This designates the boundary drawn between a number of smaller errors and an equal number of larger ones, so that it is more likely that an observation falls within any wider limit as outside it.

Solving the equation
\[
     \int_0^x d^{-t^2} dt = \int_x^\infty e^{-t^2}dt,
\]
we find that $x = 0,4769364 = h\epsilon''$, so that
\[
     \epsilon''=\alpha \times 0.8453 \epsilon' = 0.6745 \epsilon.
\]

The probability of an error smaller than $\alpha\epsilon''$ is to the probability of one larger as the value of the integral $\int e^{-t^2}dt$ from $t = 0$ to
$t= \alpha\times 0.4769364$ is the the value of the same integral from $t = \alpha\times 0,4769364$ to $t = \infty$.  From the known table of this integral we find the following for several values of $\alpha$:
\[
\begin{matrix*}[l]
\alpha   =1 &\cdots\cdots&1:1\\
\alpha   =1.25 &\cdots\cdots&1:1.505\\
\alpha   =1.5 &\cdots\cdots&1:2.209\\
\alpha   =1.75 &\cdots\cdots&1:3.204\\
\alpha   =2 &\cdots\cdots&1:4.638\\
\alpha   =3 &\cdots\cdots&1:30.51\\
\alpha   =4 &\cdots\cdots&1:142.36
\end{matrix*}
\]

\subsubsection*{The German original}

Der Erfolg der Bestimmung der endlichen Elemente aus diesen Bedingungsgleichungen, noch mehr aber die Schätzung ihrer wahrscheinlichen, aus den Beobachtungsfehlern entstehenden Unsicherheit, hängt hauptsächlich von der richtigen Würdigung der Bedingungsgleichungen ab. Es war daher nothwendig, über diesen Gegenstand eine eigene Untersuchung anzustellen, deren Resultat ich bereits seit einigen Jahren mit Vortheil benutzt habe.

Nach der von Gauss gegebenen Theorie der kleinsten Quadrate ist die Wahrscheinlichkeit, einen Fehler $\Delta$ zu begehen,
\[
        \phi(\Delta) = \frac{h}{\sqrt{\pi}} e^{-h^2\Delta^2}
\]
(\emph{Theoria mot.\ corp.\ coel.\ P.\ 212.}), wo $h$ von der Genauigkeit der Beobachtungen abhängt. Mittelst dieses Ausdrucks kann man leicht aus einer vorhandenen Reihe von Beobachtungen den wahrscheinlichen Fehler einer einzelnen bestimmen, unter der Voraussetzung, dass die wirklich vorkommenden Fehler von allen beständigen Einwirkungen frei, und nur durch die Unvollkommenheiten der Instrumente und Sinne erzeugt sind. Man hat nämlich, desto näher, je grösser die Anzahl der Beobachtungen ist, das arithmetische Mittel aus allen Fehlern, sämmtlich mit gleichem Zeichen genommen, welches wir $\epsilon$ nennen wollen,
\[
        = 2\int_0^\infty \phi(\Delta)\Delta d\Delta =\frac{1}{h\sqrt{\pi}};
\]
und auch die Quadratwurzel aus dem arithmetischen Mittel der Quadrate der Fehler, welche wir durch $\epsilon'$ bezeichnen wollen, aus der Gleichung
\[
         \epsilon'^2 = 2 \int_0^\infty \phi(\Delta) \Delta^2d\Delta = \frac{1}{2h^2}.
\]

Je zahlreicher nämlich eine vorhandene Beobachtungsreihe ist, mit desto mehr Rechte wird man annehmen können, dass die Fehler darin so vorkommen, wie es die Gausssche Theorie erfordert; das aus der Vergleichung einer sehr zahlreichen Reihe mit einer ihr so gut als möglich entsprechenden Theorie folgende $\epsilon$ oder $\epsilon'$, wird nun den wahrscheinlichen Fehler einer Beobachtung, den ich durch $\epsilon''$ bezeichnen werde, geben. Ich verstehe unter dieser Benennung die Grenze, die eine Anzahl kleinerer Fehler von einer gleichen Anzahl grösserer trennt, so dass es wahrscheinlicher ist, eine Beobachtung innerhalb jeder weiteren Grenze von der Wahrheit abirren zu sehen, als ausserhalb derselben.

Durch die Auflösung der Gleichung
\[
     \int_0^x d^{-t^2} dt = \int_x^\infty e^{-t^2}dt
\]
findet man $x = 0,4769364 = h\epsilon''$, so dass man hat
\[
     \epsilon''=\alpha \times 0.8453 \epsilon' = 0.6745 \epsilon.
\]

Die Wahrscheinlichkeit eines Fehlers, kleiner als $\alpha\epsilon''$, verhält sich zu der eines grössern, wie der Werth des Integrals $\int e^{-t^2}dt$ von $t = 0$ bis $t= \alpha\times 0.4769364$, zu dem Werthe desselben Integrals von $t = \alpha\times 0,4769364$ bis $t = \infty$ genommen. Für einige Werthe von $\alpha$ findet man, aus den bekannten Tafeln dieses Integrals:
\[
\begin{matrix*}[l]
\alpha   =1 &\cdots\cdots&1:1\\
\alpha   =1.25 &\cdots\cdots&1:1.505\\
\alpha   =1.5 &\cdots\cdots&1:2.209\\
\alpha   =1.75 &\cdots\cdots&1:3.204\\
\alpha   =2 &\cdots\cdots&1:4.638\\
\alpha   =3 &\cdots\cdots&1:30.51\\
\alpha   =4 &\cdots\cdots&1:142.36
\end{matrix*}
\]


\subsection{Thomas Galloway, 1796--1851}\label{subsec:Galloway}

The British mathematician Thomas Galloway wrote on astronomy but worked as an actuary beginning in 1833.  His \emph{Treatise on Probability} \cite[p.~144]{Galloway:1839}, published as a book in 1839, first appeared as the article on probability in the 7th edition of the Encyclopedia Britannica.  

In the preface of his \emph{Treatise} (page xi), Galloway explained that, ``In the investigation of the most probable mean value of a quantity, to be determined in magnitude or position, from a series of observations liable to error, and the determination of the limits of probable uncertainty, I have followed the very general and elegant analysis of Poisson.''  Because Poisson was much clearer than Laplace, Galloway played an important role in making the mathematics of Laplace's asymptotic theory understood in Britain.  One indication of the influence of Galloway's \emph{Treatise} is that Karl Pearson recommended it to his readers in the book on the philosophy of science that he published in 1892, before he began his research in statistics \cite[pp.~177, 180]{Pearson:1892}.

When tabulating normal probabilities, Galloway wrote $\tau$ for the number of moduli and $\Theta$ for the corresponding probability---the probability that a quantity following the error law is within $\tau$ moduli.  His table stopped at $\tau=3$; and he seems to have agreed with Fourier that this was the limit of practical certainty.  In his discussion of Bernoulli's theorem on page~144, for example, he pointed out that $\Theta$ ``approaches nearer and nearer to certainty" as $\tau$ increases, adding that ``it may be seen, by referring to the table, that it is only necessary to have $\tau=3$ in order to have $\Theta=\cdot9999779$''.

\subsection{George Biddell Airy, 1801--1892}\label{subsec:Airy}

A prolific mathematician, Airy was the British Astronomer Royal from 1835 to 1881. His book on the theory of errors, entitled \emph{On the Algebraical and Numerical Theory of Errors of Observation and the Combination of Observations} and first published in 1861, might be considered the first manual on the Gaussian theory of errors published in English.  Curiously, however, he relied on Laplace for this basic theory and did not mention Gauss's name.  He included this statement in the preface to the first edition (p.~vi):
\begin{quotation}
No novelty, I believe, of fundamental character, will be found in these pages.  At the same time I may state that the work has been written without reference to or distinct recollection of any other treatise (excepting only Laplace's \emph{Th\'eorie des Probabilit\'es}); and the methods of treating the different problems may therefore differ in some small degrees from those commonly employed.
\end{quotation}
Further editions appeared in 1875 and 1879.

On p.\ 15 of the book, Airy introduced the name \emph{modulus} for the constant $c$ in the expression
\[
                    \frac{1}{c\sqrt{\pi}} . e^{-\frac{x^2}{c^2}} . \delta x
\]
for the probability that an error falls between $x$ and $x+\delta x$.

As customary in the Gaussian tradition, Airy did not discuss practical certainty.  When discussing the law of error, on p.\ 17, he did observe that ``after the Magnitude of Error amounts to $2.0 \times \text{Modulus}$, the Frequency of Error becomes practically insensible'', but here he was referring to the density of the normal distribution, not to its tail probabilities.

\subsection{Augustin Cournot, 1801--1877}\label{subsec:Cournot}

In 1833, Cournot published a translation into French of John Herschel's \emph{Treatise on Astronomy}, which had appeared in English that same year.  In an appendix to the translation, he discussed the application of probability to astronomical observations \cite{Martin:2012}.  Here we find this statement about practical certainty \cite[vol.~XI.2, p.~686]{Cournot:1973}.
\begin{quote}
A probability of 1000 to 1 is almost considered equivalent to certainty, and one can hardly make the same judgement about a probability of 12 to 1.
\end{quote}

Already in 1828, Cournot had already began working on the ideas that became his \emph{Exposition de la th\'eorie des chances et des probabilit\'es}, but he finally completed and published the book only in 1843.%
\footnote{For Cournot 1828 article on probability, together with commentary by Bernard Bru and Thierry Martin, see pp.~442--453 of Volume XI-1 of Cournot's complete works \cite{Cournot:1973}.  See also the edition of the \emph{Exposition} that appeared in 1984, with introduction and notes by Bru, as Volume I of the complete works.}
Below I translate two passages from the \emph{Exposition}, a brief passage from \S95 discussing the agreement between Bernoullian and Bayesian methods, and an extended passage from \S111 criticizing the p-hacking of the census.  Cournot discussed p-hacking further in \S\S102 and~112--114, concluding that judgement about the meaningfulness of such observed differences is ultimately a matter of philosophical (non-numerical) probability.  His critique of p-hacking has been discussed by Bernard Bru \cite{Bru:2004} and Michel Armatte \cite{Armatte:2004,Armatte:2006}.  
 
\subsubsection*{Cournot on Bayes, from \cite[\S95]{Cournot:1843}}

When the numbers \ldots are very large, \ldots, the result from Bayes's rule no longer differs noticeably from the calculation that Bernoulli's theorem would give.  This is the way it should be, because the truth of Bernoulli's theorem is independent of any hypothesis concerning the initial choice of the urn. In this case it is not  (as many authors seem to have imagined) Bernoulli's rule that becomes exact by approximating Bayes's rule; it is Bayes's rule that becomes exact, or acquires an objective value that it did not have before, by coming together with Bernoulli's rule.

\noindent
\paragraph{The French original.}
Quand les nombres  \ldots    sont très grands, \ldots  le résultat trouvé par la régle de Bayes ne diffère plus sensiblement de calcul que donnerait le théorème de Bernoulli.  Il faut bien qu'il en sont ainsi, puisque la vérité du théorème de Bernoulli est indépendante de toute hypothèse sur le tirage préalable de l'urne.  Ce n'est point dans ce cas (comme beaucoup d'auteurs ont paru se le figurer) la régle de Bernoulli qui devient exacte en se rapprochant de la règle de Bayes; c'est la règle de Bayes qui devient exacte, ou acquiert une valeur objective qu'elle n'avait pas; en se confondant avec la règle de Bernoulli.

\subsubsection*{Cournot on p-hacking, from \cite[\S111]{Cournot:1843}}

\ldots Clearly nothing limits the number of the aspects under which we can consider the natural and social facts to which statistical research is applied nor, consequently, the number of variables according to which we can distribute them into different groups or distinct categories.  Suppose, for example, that we want to determine, on the basis of a large number of observations collected in a country like France, the chance of a masculine birth.  We know that in general it exceeds 1/2.  We can first distinguish between legitimate births and those outside marriage, and as we will find, with large numbers of observations, a very appreciable difference between the values of the ratio of masculine births to total births, depending on whether the births are legitimate or illegitimate, we will conclude with very high probability that the chance of a masculine birth in the category of legitimate births is appreciably higher than the chance of the event in the category of births outside marriage.  We can further distinguish between births in the countryside and births in the city, and we will arrive at a similar conclusion.  These two classifications come to mind so naturally that they have been an object for examination for all statisticians.

Now it is clear that we could also classify births according to their order in the family, according to the age, profession, wealth, and religion of the parents; that we could distinguish first marriages from second marriages, births in one season of the year from those in another; in a word, that we could draw from a host of circumstances incidental to the fact of the birth, of which there are indefinitely many, producing just as many groupings into categories.  It is likewise obvious that as the number of groupings thus grows without limit, it is more and more likely \emph{a priori} that merely as a result of chance at least one of the groupings will produce, for the ratio of the number of masculine births to the total number of births, values appreciably different in the two distinct categories.  Consequently, as we have already explained, for a statistician who undertakes a thorough investigation, the probability of a deviation of given size not being attributable to chance will have very different values depending on whether he has tried more or fewer groupings before coming upon the observed deviation.  As we are always assuming that he is using a large number of observations, this probability will nevertheless have an objective value in each system of groupings tried, inasmuch as it will be proportional to the number of bets that the experimenter would surely win if he repeated the same bet many times, always after trying just as many perfectly similar groupings, providing also that we had an infallible \emph{criterium} for distinguishing the cases where he is wrong from those where he is right.

But usually the groupings that the experimenter went through leave no trace; the public only sees the result that seemed to merit being brought to its attention.  Consequently, an individual unacquainted with the system of groupings that preceded the result will have absolutely no fixed rule for betting on whether the result can be attributed to chance.  There is no way to give an approximate value to the ratio of erroneous to total judgments a rule would produce, even supposing that a very large number of similar judgments were made in identical circumstances.  In a word, for an individual unacquainted with the groupings tried before the deviation $\delta$ was obtained, the probability corresponding to that deviation, which we have called $\Pi$, loses all objective substance and will necessarily carry varying significance for a given magnitude of the deviation, depending on what notion the individual has about the \emph{intrinsic importance} of the variable that served as the basis for the corresponding grouping into categories.

\subsubsection*{The French original}

\ldots  Il est clair que rien ne limite le nombre des faces sous lesquelles on peut consid\'erer les \'ev\'enements naturels ou les faits sociaux auxquels s'appliquent les recherches de statistique, ni, par suite, le nombre des caract\`eres d'apr\`es lesquels on peut les distribuer en plusieurs groupes ou cat\'egories distinctes.  Supposons, pour prendre un exemple, qu'il s'agisse de d\'eterminer, d'apr\`es un grand nombre d'observations recueillies dans un pays tel que la France, la chance d'une naissance masculine qui, en g\'en\'eral, comme on le sait, surpass 1/2:  on pourra distinguer d'abord les naissances l\'egitimes des naissances hors mariage; et comme on trouvera, en op\'erant sur de grands nombres, une diff\'erence tr\`es-sensible entre les valeurs du rapport du nombre des naissance masculines au nombre total des naissances, selon qu'il agit  d'enfants l\'egitimes ou naturels, on en conclura avec une probabilit\'e  tr\`es-grande que la chance d'une naissance masculine, dans la cat\'egoire des naissances l\'egitime, surpasse sensiblement la chance du m\^eme \'ev\'enement, dans la cat\'egoire des naissances hors mariage.  On pourra distinguer encore les naissances dans les campagnes des naissances dans les villes, et l'on arrivera \`a une conclusion analogue.  Ces deux classifications s'offrent si naturellement \`a l'esprit, qu'elles ont \'et\'e un objet d'\'epreuve pour tous les statisticiens.

Maintenant il est clair qu'on pourrait aussi classer les naissances d'apr\`es l'ordre de primog\'eniture, d'apr\`es l'\^age, la profession, la fortune, la religion des parents; qu'on pourrait distinguer les premi\`eres noces des secondes, les naissances survenues dans telle saison de l'ann\'ee, des naissances survenues dans une autre saison; en un mot, qu'on pourrait tirer d'une foule de circonstances accessoires au fait m\^eme de la naissance, des caract\`eres, en nombre ind\'efini, qui serviraient de base \`a autant de syst\`emes de distribution cat\'egorique.  Il est pareillement \'evident que, tandis que le nombre des coupes augmente ainsi sans limite, il est \emph{\`a priori} de plus en plus probable que, par le seul effet du hasard, l'une des coupes au moins offrira, pour le rapport du nombre des naissances masculines au nombre total des naissancees, dans les deux cat\'egories oppos\'ees, des valueurs sensiblement diff\'erentes.  En cons\'equence, ainsi que nous l'avons d\'ej\`a expliqu\'e, pour le statisticien qui se livre \`a un travail de d\'epouillement et de comparaison, la probabilit\'e qu'un \'ecart de grandeur donn\'ee n'est pas imputable aux anomalies du hasard, prendra des valeurs tr\`es-diff\'erentes, selon qu'il aura essay\'e un plus ou moins grand nombre de coupes avant de tomber sur l'\'ecart observ\'e.  Comme on suppose toujours qu'il a op\'er\'e sur de grands nombres, cette probabilit\'e \ldots n'en aura pas moins, dans chaque syst\`eme d'essais, une valeur objective, en ce sens qu'elle sera proportionnelle au nombre de paris que l'exp\'erimentateur gagnerait effectivement, s'il r\'ep\'etait un grand nombre de fois le m\^eme pari, toujours \`a la suite d'autant d'essais parfaitement semblables, et si l'on poss\'edait d'ailleurs un \emph{criterium} certain pour distinguer les cas o\`u il se trompe des cas o\`u il rencontre juste.

Mais ordinairement ces essais par lesquels l'exp\'erimentateur a pass\'e ne laissent pas de traces; le public ne conna\^it que le r\'esultat qui a paru m\'eriter de lui \^etre signal\'e; et en cons\'equence, une personne \'etrang\`ere au travail d'essais qui a mis ce r\'esultat en \'evidence, manquera absolument de r\`egle fixe pour parier que le r\'esultat est ou non imputable aux anomalies du hasard.  On ne saurait assigner approximativement la valeur du rapport du nombre des jugements erron\'es qu'elle portera, au nombre des jugements port\'es, m\^eme en supposant tr\`es-grand le nombre des jugements semblables, port\'es dans des circonstances identiques.  En un mot, la probabilit\'e que nous avons appel\'ee $\Pi$, et qui correspond \`a l'\'ecart $\delta$, perdra, pour la personne \'etrang\`ere aux essais qui ont manifest\'e cet \'ecart, toute consistance objective; et, selon l'id\'ee que cette personne se fera de la \emph{valueur intrins\`eque} du caract\`ere qui a servi de base \`a la division cat\'egorique correspondante, elle devra porter des jugements diff\'erents, la grandeur de l'\'ecart signal\'e restant la m\^eme.

\subsection{Jules Gavarret, 1809--1890}\label{subsec:Gavarret}

In 1840, Jules Gavarret published \emph{Principes généraux de statistique médicale}, the first book on the use of probability to evaluate medical therapies \cite{Gavarret:1840,Huth:2008}.  For Gavarret, introducing probability into medicine was a way of bringing medicine up to the level of the most exact sciences, which also, according to Laplace and Poisson, rested ultimately only on probabilities (p.~39).  On page 257, Gavarret appealed to Poisson's authority to support the choice of 2 moduli as the level of probability sufficient for practical certainty:
\begin{quotation}
But to make these formulas immediately applicable to the questions we are concerned with, we must transform them in a very simple way. To this end, recall the general principle established on page 39, namely that once an observer has arrived at a high degree of probability for the existence of a fact, he may use the fact as if he were absolutely certain of it. Let us therefore agree on a probability after which any therapeutic fact can and should be accepted without dispute.  This probability must satisfy two important conditions: one, to be sufficiently high to leave no doubt in people's minds; the other, not to require too large a number of observations in order for the ratios provided by the statistics we have collected to properly approximate the average chance we are estimating. The choice of such a probability, one that can and should satisfy us, would have been very delicate; but fortunately we can rely in this matter on an authority whose importance no one, surely, will try to dispute. When M.\ Poisson set out in his book the rules which should govern the search for possible errors in the judgements of juries,%
\footnote{This is a reference to \S135 of Poisson's book.}
the highest probability that he would give to his propositions, in order to consider himself justified in considering them as free from any reasonable objection, is:
\[
    \text{P} = 0.9953; \text{that is to say, betting odds of 212 to 1}.
\]
\end{quotation}

\begin{quotation}
Mais pour rendre ces formules immédiatement applicable aux questions dont nous nous occupens, nous devons leur faire subir une transformation très simple.  A cet effect, rappelons nous le principe général établi (pag. 39), à savoir que: du moment ou un observateur est arrivé à un haut degré de probabilité relativement à l'existence d'un fait, il peut l'employer comme s'il était absolument certain.  Convenons donc d'une probabilité à partir de laquelle tout fait thérapeutique puisse et doive être admis sans contestation.  Cette probabilité doit satisfaire à deux conditions importantes: l'une, d'être assez élevée pour ne laisser aucun doute doute dans les esprits; l'autre, de ne pas exiger un nombre trop grand d'observations, pour que les rapports fournis par la statistique recueillie se rapprochent convenablement de la chance moyenne cherchée.  Le choix de cette probabilité, dont on puisse se contenter, eut été fort délicat à faire; mais heureusement nou pouvons nous appuyer à cet égard sur une autorité dont personne sans doute n'essayera de contester l'importance.  Quand M.\ Poisson a exposé dans son ouvrages les règles qui devaient présider à la recherche des erreurs possibles dans les jugements du jury, la probabilité la plus élevée qu'il ait donné à ses propositions, pour se croire fondé à les considérer comme à l'abri de toute objection raisonnable, est:
\[
    \text{P} = 0.9953; \text{c'est à dire 212 à parier contre 1}.
\]
\end{quotation}

Today we teach students to estimate the standard error for the difference between two estimated proportions $\hat{p}_1$ and $\hat{p}_2$ by the formula
\[
             \text{standard error}\left(\hat{p}_1 - \hat{p}_2\right) = \sqrt{\frac{\hat{p}_1(1-\hat{p}_1)}{n_1} - \frac{\hat{p}_2(1-\hat{p}_2)}{n_2}},
\]
where $n_i$, for $i=1,2$, is the number of individuals from which the proportion $\hat{p}_i$ is estimated.  Gavarett used the same formula, with the additional factor $\sqrt{2}$, to estimate the modulus.  He calculated 2 moduli using this formula for a number of examples.   Here are two examples he gave on pp.~157--158:
\begin{itemize}
\item
If one treatment results in 100 deaths out of 500 patients, while a second results in 150 deaths out of 500 patients, then we have an observed difference of $0.1$, with a limit of possible errors $0.07694$.  Here Gavarret concludes:
\begin{quotation}
The difference in calculated death rates is greater than this \emph{limit of possible errors} in the \emph{a posteriori} conclusion.  So we must recognize that the first treatment really is better than the second.

La \emph{difference} entre les mortalit\'es obtenues est sup\'erieure \`a cette \emph{limite des erreurs possibles} dans la conclusion \emph{\`a posteriori}, nous devons donc reconna\^itre qu'en r\'ealit\'e la premi\`ere m\'edication est sup\'erieure \`a la seconde.
\end{quotation}
\item
If one treatment results in 100 deaths out of 500 patients, while a second results in 130 deaths out of 500 patients, then we have an observed difference of $0.06$, with a limit of possible errors $0.07508$.  Here Gavarret concludes:
\begin{quotation}
Evidently, because the difference between the two death rates is less than this \emph{limit of possible errors} in the \emph{a posteriori} conclusion, we must recognize that this variation in the results tells us nothing, and that we are not authorized to prefer one of the two methods to the other.

\'Evidemment, puisque la \emph{diff\'erence} entre les deux mortalit\'es moyennes est inf\'erieure \`a cette \emph{limite des erreurs possible} dans la conclusion \emph{\`a posteriori}, nous devons reconna\^itre, que cette variation dans les r\'esultats ne nous enseigne rien, que nous ne sommes pas autoris\'es \`a pr\'ef\'erer une des deux m\'ethodes \`a l'autre..
\end{quotation}
\end{itemize}

\subsection{Wilhelm Lexis, 1837--1914}\label{subsec:Lexis}

A prominent German economist and statistician, Lexis published his introduction to population statistics \emph{Einleitung in die Theorie der Bev\"olkerungsstatistik} \cite{Lexis:1875}, in 1875.  

On p.~98, he gave this small table for normal probabilities, $F_u$ being the probability of a quantity estimated being within $u$ moduli of the estimate.  
\begin{center}
\begin{tabular}{l|l||l|l}
$u$ & $F_u$ & $u$ & $F_u$\\
0,10 & 0,11246 & 1,50 & 0,966105\\
0,20 & 0,22270 & 2,00 & 0,995322\\
0,30 & 0,32863 & 2,50 & 0,999593\\
0,40 & 0,42839 & 3,00 & 0,999977909\\
0,50 & 0,52050 & 4,00 & 0,999999985\\
1,00 & 0,84270 & 5,00 & 0,999999999998
\end{tabular}
\end{center}

Lexis consistently used Fourier's criterion of three moduli for practical certainty. This passage, from p.~100, is typical of the explanations he gives for choosing $u$ to be $3$:
\begin{quote}
For the purposes of statistics it should however be more appropriate to take $u$ so large that $F_u$ comes very near to one and therefore expresses a probability that can be considered in practice equal to certainty.  It suffices, as before, to set $u$ equal to $3$, and we then obtain the probability $F_3 = 0,999978$ for the limit equation \ldots.  
\end{quote}

\emph{In the original:}  
\begin{quote}
F\"ur die Zwecke der Statistik d\"urfte es jedoch geeigneter sein, f\"ur $u$ eine so grosse Zahl zu nehmen, dass $F_u$ der Einheit sehr nahe kommt, also eine Wahrscheinlicbkeit ausdr\"uckt, die in der Praxis der Gewissheit gleich geachtet werden kann. Es gen\"ugt, wie oben, $u$ gleich $3$ zu setzen, und man erhalt alsdann die Wahrscheinlichkeit $F_3 = 0,999978$ f\"ur die Grenzgleichung \ldots.)
\end{quote}

\subsection{Francis Edgeworth, 1845--1926}\label{subsec:Edgeworth}

As noted in \S\ref{subsec:Edge}, Edgeworth introduced the English word \emph{significant} into statistical testing in a paper read at the Jubilee meeting of the Statistical Society of London in 1885 \cite{Edgeworth:1885}.  In this paper, Edgeworth refers the reader to Quetelet, Galton, and Jevons for details on the law of error, but he uses the modulus and Fourier's criterion, repeated by Lexis, of thrice the modulus.  Here are a few key quotations.
\begin{quote}
From p.\ 182:  The science of Means comprises two main problems: 1.\ To find how far the difference between any proposed Means is accidental or indicative of a law?  2.\ To find what is the best kind of Mean; whether for the purpose contemplated by the first problem, the elimination of chance, or other purposes? \ldots  The first problem investigates how far the difference between the average above stated and the results usually obtained in similar experience where pure chance reigns is a significant difference; indicative of the working of a law other than chance, or merely accidental. \ldots
\end{quote}
\begin{quote}
\ldots out of a set of (say) N statistical numbers which fulfil the law of error, we take one at random, it is exceedingly improbable that it will differ from the Mean to the extent of twice, and \emph{\`a fortiori} thrice, the modulus.
\end{quote}
\begin{quote}
From p.\ 188:  \ldots we shall find that the observed difference between the proposed Means, namely about 2 (inches) far exceeds thrice the modulus of that curve, namely 0*2. The difference therefore ``comes by cause.''
\end{quote}
In his report on the discussion of the paper (p.~217), the president of the session reported that when pressed by the Italian statistician Luigi Perozzo on whether his paper contained anything new, Edgeworth had said that ``he did not know that he had offered any new remarks, but perhaps they would be new to some readers.  He had borrowed a great deal from Professor Lexis.''

Edgeworth again used \emph{significant} in his article on probability in the 11th edition of the \emph{Encyclopedia Britannica} \cite[\S137]{Edgeworth:1911}.  There he explained that the method discussed in his 1885 paper was a way of deciding whether a difference is real without resorting to a complete inverse (Bayesian) analysis.
\begin{quote}
This application of probabilities not to the actual data but to a selected part thereof, this economy of the inverse method, is widely practised in miscellaneous statistics, where the object is to determine whether the discrepancy between two sets of observation is accidental or significant of a real difference.
\end{quote} 

The following passage appears in Edgeworth's 1887 article on the rejection of discordant observations \cite[pp.~369--370]{Edgeworth:1887}:
\begin{quote}
There is something paradoxical in Cournot's proposition that a certain deviation from the Mean in the case of Departmental returns of the proportion between male and female births is significant and indicative of a difference in kind, provided that we select at random a single French Department; but that the same deviation may be accidental if it is the maximum of the respective returns for several Departments.  There is something plausible in De Morgan's implied assertion that the deficiency of seven in the first 608 digits of the constant $\pi$ is theoretically not accidental; because the deviation from the Mean 61 amounts to twice the Modulus of that probability curve which represents the frequency of deviation for any assigned digit. I submit, however, that Cournot is right, and that De Morgan, if he is serious in the passage referred to, has committed a slight inadvertence. When we select out of the ten digits the one whose deviation from the Mean is greatest, we ought to estimate the improbability of this deviation occurring by accident, not with De Morgan as $1 - \theta(1\cdot63)$, corresponding to odds of about 45 to 1 against the observed event having occurred by accident; but as $1-\theta^{10}(1\cdot63)$, corresponding to odds of about 5 to 1 against an accidental origination.
\end{quote}

\subsection{Mansfield Merriman, 1848--1925}\label{subsec:Merriman}

A prominent American mathematician and engineer, Merriman taught at Yale and Lehigh and worked at different times in his country's Corps of Engineers and Coast and Geodetic Survey.  He enters our story as an authority on Gauss's theory of least squares. In one of his earliest publications on the subject, he wrote \cite[p.~195]{Merriman:1874}:
\begin{quote}
To Gauss is also due the development of the algorithm of the method, the formulae for probable error, the determination of weights, the method of correlatives, and many other features of the subject, as well as numerous practical applications with which his writings abound. Very few branches. of science owe so large a proportion of subject matter to the labours of one man. 
\end{quote}

\subsection{Arthur Schuster, 1851--1934}\label{subsec:Schuster}

Schuster was born in Germany but in 1870 he followed his parents to England, where he became a prominent physicist.  He is best remembered for coining the term the term \emph{periodogram} and analyzing it statistically.

Schuster introduced the word \emph{periodogram} in an 1898 article on the evidence for a 26-day cycle in the weather.  In this article \cite[p.~18]{Schuster:1898}, he described an intensity that would be exceeded only one time in 23 as one that would not ``justify us \ldots to consider a real periodicity as proved, although we might be encouraged to continue the investigation by taking an increased number of events into account.''  In a 1906 article on the periodicity of sunspots \cite[p.~79]{Schuster:1906}, he was more exigent:
\begin{quote}
\ldots
The probability of an intensity greater than $h$ times the average value is $e^h$, and we may perhaps begin to suspect a real periodicity when this value is 1 in 200. This gives 5$\cdot$3 as the value of $h$ and 80,000 as the smallest value of the intensity which invites further discussion. When $h$ has the value 8, the probability of an intensity greater than h times the expectancy is 1 in 3,000 and we may begin to be more confident that there is some definite cause at work to bring up the periodogram to that value. The intensity in that case is 120,000. When $h$ is 16, the chances of being misled by accident is only one in a million.
\end{quote}

\subsection{Karl Pearson, 1857--1936}\label{subsec:Pearson}

Karl Pearson is remembered as the driving force behind the British school of biometry at the beginning of the 20th century.  One of his roles was editor, from its founding in 1901 until his death in 1936, of the journal \emph{Biometrika}.  The early issues of the journal provide a convenient view on how he and his followers talked about statistical testing at the beginning of the century.

In \emph{Biometrika}'s first volume, we find ``perhaps significant'', ``more probably significant'', and ``certainly significant''.  Here are some additional instances of the Edgeworthian ``significant'':
\begin{itemize}
\item
In the very first issue, from Pearson's close collaborator W.\ F.\ R.\ Weldon  \cite[p.~119]{Weldon:1901}:  ``With probable errors of the order indicated by Tables I. and II., it is unlikely that any of these differences are significant. Even in the case of the last pair of entries the difference, although it is considerable ($0\cdot0229$ mm.), is less than twice the probable error of the determination.''
\item
In the second issue, from Oswald H.\ Latter \cite[p.~167]{Latter:1902}:  ``To test whether any deviation is significant, $M_r$ is taken as the mean of the whole race of Cuckoos and $M_s$ the mean of Cuckoo's eggs found in the nest of any one species of foster-parent: the standard deviation ($\sigma_s$) of such eggs is also ascertainied. The value of $M_r - M_s$ is then compared with that of $0\cdot 67449 \sqrt{\frac{\sigma_r^2}{n_1}+\frac{\sigma_s^2}{n_2}}$, where $n_1 =$ total number of Cuckoo's eggs and $n_2 =$ the number of Cuckoo's eggs in the nests of the species in question, which is the probable error of $M_r - M_s$ due to random sampling. If the value of $M_r - M_s$ be not at least $1.5$ to $3$ times as great as the value of the other expression the difference of $M_r$ and $M_s$ is not definitely significant.''
\item
 In volume 7, for 1909/1910, the American James Arthur Harris (1880--1930) wrote ``\ldots I follow the rather common example of statisticians in regarding differences of at least $2\cdot5$ times their probable errors as significant'' \cite[p.~458]{Harris:1909/1910}.
 \item
 In a 1912 article co-authored by Pearson himself \cite[p.~301]{Benington/Pearson:1912}: ``Hence the difference is more than three times the probable error and likely to be significant.''
\end{itemize}
These quotations indicate that Pearson and his school consistently used \emph{significant} in the Edgeworthian sense, and that they still measured the likelihood of significance with the probable error rather than the standard error.  A difference of more than three probable errors was judged definitely significant, a difference of less than two was thought unlikely to be significant.

In later years, however, we see some non-Edgeworthian uses of \emph{significance} creep into \emph{Biometrika}.  Here are some examples.
\begin{itemize}
\item
In the volume for 1908/1909, J.\ F.\ Tocher \cite[p.~163]{Tocher:1908/1909}, writes ``it is possible that a locality may exhibit a difference or differences almost or just significant for one or more colour classes\ldots''.
\item
In the volume for 1914/1915, in an article on the variate difference method co-authored by Pearson himself \cite[p.~347]{Cave/Pearson:1914/1915}:  ``Stripped therefore of the common time factor the \emph{Synthetic Index} will be seen to be no very appropriate measure of trade, business activity, and spare money for savings and luxuries. With \emph{Post, Stamp Duties and Savings}, it has probably only a spurious relationship, expenditure on railways has little influence, that on luxuries is very slightly significant, or indeed in the case of tobacco negative.''
\item
In the volume for 1918/1919, in an article on psychophysics Godfrey H. Thomson \cite[p.~219]{Thomson:1918/1919}:  ``The difference is therefore three times its probable error and is just significant.''
\end{itemize}

The subtle nuances of Edgeworth's \emph{significant} were definitively lost in the 1920s.  Perhaps they were too subtle to survive.  But they did survive long enough for the word to become embedded in mathematical statistics, with all its confusing awkwardness and stubborn permanence.


\subsection{Gilbert Walker, 1868--1958}\label{subsec:Walker}

Walker was already an accomplished applied mathematician when he accepted an appointment to the British meteorological office in India.  By 1914 \cite{Walker:1914}, he had published a memoir under that office's auspices deploring multiple testing in the statistical interpretation of Schuster's periodograms.  This publication may have escaped the notice of his colleagues back in Britain, but he made his point well known in a letter to the editor of \emph{Nature} in 1922 \cite[p.~511]{Walker:1922} and in an article in the \emph{Quarterly Journal of the Royal Meteorological Society} in 1925 \cite{Walker:1925}.

\paragraph{Walker's 1922 letter to the Editor of \emph{Nature}, entitled ``On periodicities'':}

T\textsc{HE} recent paper by Sir William Beveridge on ``Wheat Prices and Rainfall'' (Journal of the Royal Statistical Society, vol.\ 85, pp. 412--478, 1922) raises a rather important question of principle which is involved not only in discussions over the existence of periodicities, but also over relationships between different variables.

Before Schuster's papers on the periodogram it was customary for a period to be accepted as real provided that it had an amplitude comparable with that of the original figures under analysis; and he revolutionised the treatment of the subject by showing that if the squares of the intensities of the various periodic terms are plotted in a periodogram, and if the data are those of an entirely chance distribution, then the average value of an ordinate being $a$, the probability that a particular ordinate will equal or exceed $ka$ is $e^{-k}$. Sir William Beveridge is accordingly perfectly justified in taking Schuster's sunspot period of 11$\cdot$125 years, or Brückner's 34$\cdot$8 year period, and deciding that these periods probably occur in his wheat prices if the corresponding intensities are three or four times the average. But he, like many other investigators, goes a stage further, and after picking out the largest from a large number of intensities he applies the same criterion as if no selection had occurred. It is, how ever, clear that if we have a hundred intensities the average of which, a, is derived from a number of random figures, then the probable value of the largest of these chance intensities will not be a but will be considerably greater, and it is only when the largest amplitude actually derived materially exceeds the theoretical chance value thus obtained that reality can be inferred.

Taking the periodicities of wheat prices on pp.\ 457--459 between 5 years and 40 years,%
\footnote{Footnote by Walker:  Sir William Beveridge points out on pp.\ 423--424 that amplitudes for periods of less than 5 years are inevitably diminished, while those above 31 are diminished by the process employed for eliminating secular trend: I calculate that the intensity at 35 years should be multiplied by $(0\cdot87)^{-2}$ or 1$\cdot$3, and that at 54 by 3$\cdot$8.}
I estimate that the ``width of a line'' ranges from 0$\cdot$1 year for a 5 years' period, through 0$\cdot$5 at 12 years to 4 years at 33 years; and accordingly that the number of independent periods between 5 years and 40 is in this case about 51. The value of $a$, the average intensity, being 5$\cdot$898, it is easily seen that the chance of all the 51 random intensities being less than $3a$ is $(1 - e^{-3})^{51}$, or 0$\cdot$074, so that the chance of at least one intensity greater than $3a$ is 0$\cdot$926, not $e^{-3}$ or $0.050$, as is habitually assumed. Instead of the chance of an occurrence of $3a$ ``making a \emph{prima facie} case for enquiry'' (p.\ 424), the odds are 12 to 1 in favour of its production by mere chance. The chance of at least two intensities above $3a$ is 0$\cdot$728, of three it is 0$\cdot$470, of four 0$\cdot$248, of five 0$\cdot$109, of six 0$\cdot$0403, of seven 0$\cdot$0127, of nine 0$\cdot$00085, and of eleven 0$\cdot$00003. Thus it is not until six intensities over $3a$ are found that the chance of production by pure luck is less than 1 in 20. It is also easily found that if the chance of all the 51 intensities being less than $na$ is to be 19/20, $n$ is 6$\cdot$9; i.e. the greatest intensity for wheat price fluctuations must be 41, not 18, before the probability of its being due to luck is reduced to 1/20; and if the likelihood is to be 1/100 we must have $n = 8\cdot5$, the corresponding wheat-price intensity being 50.  Of intensities greater than 41 Sir William Beveridge  found four, and greater than 50 only two.

At first sight it might seem that the agreement between Sir William Beveridge's forecasted synthesis rainfall curve and the actual rainfall was too great to be explained by a few harmonic terms; but the correlation co-efficient of $0\cdot38$ (see p.\ 475) indicates that while $0\cdot38$ of the rainfall variations are accounted for, only $(0\cdot38)^2$, or about a seventh, of the independent factors which control these variations have been ascertained.

As pointed out in a paper ``On the Criterion for the Reality of Relationships or Periodicities,'' in the Indian Meteorological Memoirs (vol.\ 21, No.\ 9, 1914), the same principle is valid when discussing relationships. If we are examining the effect of rainfall on temperature and ascertain that the correlation coefficient between the rainfall and temperature of the same month in a particular English county is four times the probable error, we may infer that the effect is highly probable. But if we work out the co-efficients of that temperature with a hundred factors taken at random, e.g. with the monthly rainfall of Tashkend 5$\cdot$8 years previously, and pick out the largest co-efficient, it would be wrong to compare it with the average co-efficient produced by mere chance; as shown in the paper referred to, the probable value of the largest of 100 co-efficients is
$4\cdot01$  times as great as the probable value of one taken at random. 

GILBERT T. WALKER.  

Meteorological Office, Simla, August 24.

\subsection{Arthur Bowley, 1869--1957}\label{subsec:Bowley}

Bowley was a professor of economics at the London School of Economics.  Like his fellow economist Edgeworth, he was not part of Pearson's biometric circle.  He published several textbooks on statistics, beginning with the first edition of his \emph{Elements of Statistics} in 1901 \cite{Bowley:1901}, where he acknowledged a debt to Edgeworth, both for Edgeworth's publications and for personal instruction.  As we see on page~6 of the book, he adopted Edgeworth's criterion of 3 moduli for practical certainty:
\begin{quote}
Without the aid of statistical method, the averages obtained show mere numbers from which no logical deductions can be made. With the help of this knowledge, it can be seen whether the change from year to year is significant or accidental; whether the figures show a progressive or periodic change; whether they obey any law or not.
\end{quote}
On page~313, he cites Edgeworth \cite{Edgeworth:1885} as authority for the proposition that an apparent difference of 3 moduli signifies a real difference.
\begin{quote}
\ldots the modulus of a difference is most useful in comparing two groups selected as having certain qualities. Thus Professor Edgeworth discusses whether an ascertained difference of 2 inches between the average heights of a large number of criminals and that of the general population is significant; and finding that the modulus for the difference between two random groups is only 0.08, holds that there is a cause of the difference in the method of selection; that is, that criminality and low stature are found together.  We might apply the same principle to the investigation of the existence of a period in any figures; for if the modulus of the figures was $c$, the modulus for the difference between the averages of two random samples of $20$ months each would be $c\sqrt{\frac{1}{20} + \frac{1}{20}}$; if the difference between the averages of the figures for 20 Decembers and 20 Junes was 3 times this quantity the existence of a period would be established.
\end{quote}

\subsection{George Udny Yule, 1871--1951}\label{subsec:Yule}

After studying mathematical physics, Yule became a statistician as an assistant to Pearson.  He later quarreled with Pearson and went his own way on a number of points.  In 1897 \cite{Yule:1897}, he introduced the name \emph{standard error} for the standard deviation of an estimate.  The first edition of his \emph{Theory of Statistics} \cite{Yule:1911} appeared in 1911.  

On page~262 of the first edition, we find this nod to Edgeworth's \emph{significant}:
\begin{quote}
\ldots if we observe a different proportion in one sample from that which we have observed in another, the question again arises whether this difference may be due to fluctuations of simple sampling alone, or whether it indicates a difference between the conditions subsisting in the universes from which the two samples were drawn: in the latter case the difference is often said to be \textbf{significant}. These questions can be answered, though only more or less roughly at present, by comparing the observed difference with the standard-deviation of simple sampling. We know roughly that the great bulk at least of the fluctuations of sampling lie within a range of $\pm$ three times the standard-deviation; and if an observed difference from a theoretical result greatly exceeds these limits it cannot be ascribed to a fluctuation of ``simple sampling'' as defined in \S8: it may therefore be significant. The ``standard-deviation of simple sampling'' being the basis of all such work, it is convenient to refer to it by a shorter name. The observed proportions of A's in given samples being regarded as differing by larger or smaller errors from the true proportion in a very large sample from the same material, the ``standard-deviation of simple sampling'' may be regarded as a measure of the magnitude of such errors, and may be called accordingly the \textbf{standard error}.
\end{quote}
Yule never or hardly ever used the Edgeworthian \emph{significant}, however.  We find the word repeatedly in the book, but usually merely to mean ``important''.

\subsection{Francis J. W. Whipple, 1876--1943}\label{subsec:Whipple}

The following passage appears on p.~336 of the discussion of Brunt's paper on the Greenwich temperature records.
\begin{quotation}
Mr.\ F.\ J.\ W.\ WHIPPLE called attention to a difficulty which was not always faced when the ``reality'' of periods was under discussion. In Capt\. Brunt's paper it was stated that if the amplitude of one of the harmonic components was as great as $06\circ\cdot4$ F, the odds were ``19 to 1 in favour of the reality'' of the corresponding period, and yet the author arrived at the conclusion that such periods had little physical significance. The difficulty lay in the enunciation of the statement with regard to the probability of the occurrence of an amplitude of a specified magnitude. It might be illustrated by an example from the card-table: the chances of three aces occurring in a particular hand was very small, but the chance of three aces occurring in some hand in the course of an evening's play was very high. In the case under consideration, if a particular period was named in advance and the corresponding amplitude turned out to be large, then the probability that this was not merely casual would be high; but if all possible periods were investigated then there was good reason to expect that some of the computed amplitudes would be large.  It was stated that the standard (root-mean-square) value of the amplitudes computed from the author’s data, regarded as distributed by chance, would be $2\cdot3$ units, and as a matter of fact the sum of the squares of the amplitudes of the first forty-nine harmonics was given as 380, so that the root-mean-square was $2\cdot8$, not much in excess of the $2\cdot3$. Accordingly there was little reason to suppose that the periods were significant. Any agreement between the periods found for one element and another, for example. for temperatures at different places, would indicate a correlation between the elements and would merit further investigation. Periodogram analysis might prove a useful though laborious method for discovering such correlations.
\end{quotation}

\subsection{William Beveridge, 1879--1963}\label{subsec:Beveridge}

When Beveridge published his harmonic analysis of hundreds of years of wheat prices in 1922, he was already well known because of his work in the British civil service before and during the First World War.  In 1919, he had become director of the London School of Economics.  During the Second World War, he led a committee that produced the \emph{Beveridge Report}, which became a blueprint for Britain's postwar welfare system.

Beveridge's response to Gilbert Walker's 1922 letter to the Editor of \emph{Nature} was printed by the journal immediately after the letter.

\paragraph{Beveridge's response.}

DR.\ WALKER's note contains, I think, a valid and valuable criticism of the procedure commonly adopted hitherto in comparing individual intensities with the average intensity in harmonic analysis. It would lead me, now to modify in several ways my general discussion of the ``test of intensity'' (pp.~412--424 of my paper in the Journal of the Royal Statistical Society). I was particularly careful, however, in that paper to avoid laying stress on intensity as such. The net result of Dr. Walker's calculations is not to weaken but to confirm my main thesis: that a number of real periodicities exist in European wheat prices from 1550 to 1850.

According to these calculations, the chance of my getting by pure luck between five and forty years one intensity as great as $3a$ is $0\cdot926$, but the chance of my getting seven such intensities is $0\cdot0127$, and that of getting eleven is $0\cdot00003$. Actually I have, between five and forty years, fifteen intensities above $3a (= 17\cdot69)$; the odds are therefore $80$ to $1$ that at least nine of these intensities, and $33,000$ to $1$ that at least five of them, are not due to luck. Obviously every such intensity does, in the circumstances, present a \emph{prima facie} case for further inquiry, the object of the inquiry being to determine which of the 15 intensities have the strongest probabilities of being due to real periods.

In that inquiry the actual height of the intensity in any case (the ``test of intensity'') is only one and not necessarily the most important point for consideration. As Dr.\ Walker shows, an intensity in my periodogram of nearly seven times the average might well be due to pure luck (the odds being only 20 to 1 against it). On the other hand, a much lower intensity might represent a true and perfectly regular but weak periodicity, just as a quite small correlation co-efficient may prove a real though weak connexion, if the number of cases compared is very large. Indication of the same period in each half of a sequence when analysed separately (the ``test of continuity'') and in independent sequences (the ``test of agreement with other records'') are often more important criteria of reality than is the height of the intensity itself. The former test, at least, should never be neglected; it has led me to relegate to my fourth class as merely ``possible,'' several periods, such as those near 11, 17, and 24 years, indicated by high intensities in the whole sequence, but failing in either the first or the second half.

Ultimately, of my fifteen intensities between 5 and 40 years, I have treated only nine (at $5\cdot100$, $5\cdot671$, $5\cdot960$, $8\cdot050$, $9\cdot750$, $12\cdot840$, $15\cdot225$, $19\cdot900$, and $35\cdot500$ years respectively) as certainly or probably due to real periodicities, because they show in all cases perfect or fair continuity and in most an agreement with other records. The smallest of these fifteen intensities ($21\cdot72$ at $7\cdot417$ years) in fact equals not $3a$ but $3\cdot683a$. If with this revised figure, the probabilities are calculated in the way suggested by Dr.\ Walker, the odds that at least nine of the fifteen intensities are not due to luck work out at more than
$2000$ to $1$, while the odds in favour of seven at least are $14\cdot000$ to $1$.

This remarkable result, which seems to establish beyond all reasonable doubt the reign of periodicities in wheat prices, is not affected by the fact that of the fifteen intensities only four are so high that any one of the four, if it occurred alone and had to be judged by height alone, would have odds of more than 20 to 1 in its favour. Each intensity does not occur alone. Every period, moreover, to which I attach importance rests on more evidence than mere height in my periodogram.

With reference to the last paragraph but one of Dr.\ Walker's note, on the relation of my synthetic curve and the rainfall, I should like to emphasise the point made in my paper (pp.~449--450) that the synthetic curve as now drawn represents only a first approximation of the roughest possible character; the correlation co-efficient of $0\cdot38$ between it and the rainfall from 1850 to 1921 is sufficient to demonstrate some connexion between the wheat price cycles and the rainfall, but is in no sense to be treated as a measure of the degree of connexion. In constructing the synthetic curve, for instance, the periodicities have all been treated as of equal importance; inspection shows that weighting according to the intensities would almost certainly give a better fit and so a higher co-efficient of correlation. In many other ways a more accurate determination of the cycles is required. How high a correlation might ultimately be obtained as the result of this, it is impossible now to say, but it might easily prove to be very high indeed. Unfortunately, I have no resources for carrying my own investigations further for the present; I can only hope that others may be better placed. 

W.\ H.\ BEVERIDGE.

\paragraph{Aftermath.}

By the 1940s, British mathematicians had reached a consensus that the cycles detected by harmonic analysis of time series had little meaning or value for prediction.  An important marker was an extensive paper read to the Royal Statistical Society in 1945 by Maurice G.\ Kendall.  Kendall concluded the discussion of his paper by saying that, ``the reason why people continually discover cycles in all kinds of time series, is that they are looking for them'' \cite[pp.140--141]{Kendall:1945}.  In another discussion of cycles at the Royal Statistical Society in 1946, we see this report on a comment by Harold Jeffreys (\emph{Supplement to the Journal of the Royal Statistical Society}, Vol.\ 8, No.\ 1, p.~90): ``DR. HAROLD JEFFREYS said that he had no experience in detecting empirical periodicities in geophysical data. He had a good deal of experience of failing to find evidence for them.''

\subsection{Raymond Pearl, 1879--1940}\label{subsec:Pearl}

The American biologist Raymond Pearl, who spent most of his career at Johns Hopkins, studied with Karl Pearson for a year in 1906.  Like Yule and many other of Pearson's disciples, he eventually quarreled with the master, but he fondly acknowledged Pearson in the preface to the textbook he published 1923, \emph{Introduction to Medical Biometry and Statistics} \cite{Pearl:1923}.

Perusing the occurrences of \emph{significant} in this textbook, we might conclude that Pearl has studied and learned the Edgeworthian way of using the word but does not quite find it natural.  It is, he tells us a conventional way of talking:

On page 214:
\begin{quotation}
\noindent
\ldots Is a difference six times its probable error likely to arise from chance alone, or does it represent a really significant difference?

There has grown up a certain conventional way of interpreting probable errors, which is accepted by many workers. It has been practically a universal custom among biometric workers to say that a difference (or a constant) which is smaller than twice its probable error is probably not significant, whereas a difference
(or constant) which is three or more times its probable error is either "certainly," or at least "almost certainly," significant.
\end{quotation}

On page 217:  
\begin{quotation}
From this table it is seen that a deviation of four times the probable error will arise by chance less often than once in a hundred trials. When one gets a difference as great or greater than this he may conclude with reasonable certainty that it did not arise by chance alone, but has significant meaning. 
\end{quotation}


If we want to quibble, we can argue that Pearl has not mastered the jargon perfectly.  The antecedent of ``it'' in the first quoted sentence is the difference six times its probable error.  Edgeworth would say that this observed difference \emph{probably is} a significant difference, not that it \emph{represents} one.

\subsection{Sergei Bernstein, 1880--1968}\label{subsec:Bernstein}

Born in Odessa, Bernstein%
\footnote{In Russian, \begin{cyr}С. Н. Бернштейн\end{cyr}.  Modern transliterations render \begin{cyr}Бернштейн\end{cyr} as \emph{Bernshtein}.  But he is usually \emph{Bernstein} in the mathematical literature.}
 studied in Paris and G\"ottingen.  His dissertation, submitted to the Sorbonne in Paris in 1904, solved Hilbert's 19th problem.  He taught at the university at Kharkov from 1907 until 1932, when he moved to the Academy of Sciences in Leningrad; later he taught in Moscow.  

In 1927, Bernstein published his course on probability as a book \cite{Bernstein:1927}.   We can call the book Laplacean tradition, for it discusses practical certainty, and it often appeals to Laplace's theorem.  Bernstein mentions Lexis, Charlier, and Bowley but relies heavily on the authority of Andrei Markov, whose own probability course was published in Russian in 1900, with further Russian editions in 1908, 1913, and 1924 and a German edition in 1912.   

Bernstein's terminology for practical certainty influenced later developments.  The Russian word  \begin{cyr}уверенный\end{cyr} can be translated into English as \emph{confident} or \emph{sure}, and Bernstein frequently uses \begin{cyr}практически уверенный\end{cyr} for \emph{practically certain}.  On the first page of his book, he contrasts \begin{cyr}уверенность\end{cyr} (confidence) based on probabilities with \begin{cyr}абсолютная достоверность\end{cyr} (absolute certainty).  On p.~233, he gives a prediction interval for a binomial outcome as follows:
\begin{quotation}
\ldots \begin{cyr}уже при\end{cyr} $t=4$, $2\Phi(t)=0.999936$, \begin{cyr}и соответствующее неравенство\end{cyr}
\[
     \left\vert  \frac{m-np}{\sqrt{npq}} \right\vert < 4
\]
\begin{cyr}обычно считают практически достоверным. Отсюда следует, напри- 
мер, что, если мы произведем 2 500 бросаний монеты, то можно 
быть практически уверенным, что число т появлений „орла" будет 
удовлетворять неравенству 
\end{cyr}
\[
  \left\vert m - 1250\right\vert < 4\times25,
\]
\begin{cyr}т.-е.\end{cyr}
\[
  1150 < m < 1350;
\]
\end{quotation}
In English:
\begin{quotation}
\ldots already for $t=4$, $2\Phi(t)=0.999936$, and the corresponding inequality
\[
     \left\vert  \frac{m-np}{\sqrt{npq}} \right\vert < 4
\]
is usually considered practically certain. It follows, for example, that if we make 2,500 coin tosses, then we can be almost sure that the number $m$ of heads will  satisfy the inequality
\[
  \left\vert m - 1250\right\vert < 4 \times 25,
\]
or
\[
  1150 < m < 1350;
\]
\end{quotation}
Here I have translated \begin{cyr}практически уверенным\end{cyr} as \emph{almost sure}.

On p.~274, we find similar language in the description of what we now call a confidence interval:
\begin{quote}
\ldots \begin{cyr}благодаря тому, что\end{cyr} $p-p'$ \begin{cyr}следует закону Гаусса, после определения\end{cyr} $p'=\frac{m}{n}$ \begin{cyr}можно быть практически уверенным, что\end{cyr}
\[
              p' - 4\sigma < p < p' + 4\sigma;
\]
\end{quote}
In English
\begin{quote}
\ldots because $p-p'$ follows Gauss's law, after observing $p'=\frac{m}{n}$ you can be practically certain that
\[
              p' - 4\sigma < p < p' + 4\sigma;
\]
\end{quote}
Here, for the sake of variety, I have translated \begin{cyr}практически уверенным\end{cyr} as \emph{practically certain}.

Bernstein was Bayesian, but he does not draw the Bayesian/Bernoullian contrast in this elementary book, and he sounds Bernollian when using Laplace's theorem.  Jerzy Neyman attended Bernstein's probability course at Kharkov in 1915--1916, and as this was his main training in probability as a student, it may be unsurprising that he was somewhat uncertain about the role of prior probabilities when he began working in theoretical statistics and that he used the Polish \emph{ufno\'s\'n} and the English \emph{confidence} when he began calculating intervals in applied statistics; see~\cite[p.~43]{Lehmann:2011} and \S\ref{subsec:Neyman}.

\subsection{Truman L. Kelley, 1884--1961}\label{subsec:Kelley}

The following passage is drawn from pp.~102--103 of Kelley's 1924 book \emph{Statistical Method} \cite{Kelley:1923}.

\subsubsection*{Kelley's words}

The normal curve assists in establishing the degree of confidence which may be placed in statistical findings. The significance of any measure is to be judged by comparison with its probable error. If a child makes a score of 80 on a certain test and if the probable error of the score is 5, we may estimate the chances of the child's true ability being as much as 100. We assume that the distribution of the child's performances would follow a normal curve. Note that the assumption is not that the talents of children in general follow a normal distribution. This latter might be less reasonable than the one we are called upon to make. Moreover, so little difference in probabilities, except for extreme deviates, is ordinarily consequent to differences in forms of distribution, that the assumption of normality is little likely to result in serious error for such problems as the present one. For extreme deviates it generally does not matter so far as any practical deductions are concerned whether the chances are 1 in 1000 or ten times as great. Nor for smaller deviates does it make any particular difference whether the chances are 400 in 1000 or 410 in 1000. Should such differences as mentioned be significant in any particular problem, no assumption should be made, but the type of the curve should be experimentally determined.

For the problem in hand: If the P.\ E .\ is 5 the standard error is $\left(\frac{5}{.6745}\right) = 7.413$. The difference between the scores that we are concerned with is $(100-80) = 20$, which is $\left(\frac{20}{7.413}\right) = 2.698$ standard errors. The K-W Table, or more conveniently for this problem Sheppard's Tables, may be used to find the area in the tail below the point which is 2.698 standard deviations below the mean. The tables give .0035. To interpret this we should postulate the person's true ability as being 100 and his various performances distributing themselves in a normal distribution, with standard deviation equal to 7.413 around this mean. Then .0035 of the area of the curve will lie below the point 80. Accordingly if his true ability is 100, only 35 times in 10000, or 3.5 times in 1000, would a score as low or lower than 80 be expected. With such figures a person could accept the proposition that the child's ability was not as great as 100 with about as much certainty as he can start across a business street expecting not to be hit by an automobile. It is, in other words, just such a conclusion as one is justified in acting upon.

\subsection{David Brunt, 1886--1965}\label{subsec:Brunt}

In 1917,  the Welsh meteorologist David Brunt published a book on the theory of errors, \emph{The Combination of Observations} \cite{Brunt:1917}, which included a chapter on the periodogram.  The book was squarely in the Gaussian tradition, the book did not mention Laplace and did not set a standard for practical certainty.  

Brunt explained Schuster's probabilistic treatment of the Fourier coefficient, giving the following table and explanation on p.~200:
\begin{quotation}
\begin{center}
\begin{tabular}{|c|c||c|c|}
\hline
 $\kappa$ & $e^{-\kappa}$ & $\kappa$ & $e^{-\kappa}$\\
 \hline
 1     &  $\cdot3679$   &  6      &  $2\cdot4 \times 10^{-3}$\\
 2     &  $\cdot1353$   &  8      &  $3\cdot35 \times 10^{-4}$\\
 3     &  $\cdot0498$   &  10    &  $4\cdot54 \times 10^{-5}$\\
 4     &  $\cdot0183$   &  12    &  $6\cdot14 \times 10^{-6}$\\
 5     &  $\cdot00674$ &  16    &  $1\cdot13 \times 10^{-7}$\\
 \hline
 \end{tabular}
 \end{center}
 
 \smallskip

This table may be interpreted thus:---The chance of obtaining for the square of a Fourier coefficient a value greater than three times its expectancy or mean value is $\cdot$0498, or about 1 in 20. So that, if on analysing a series of observations we obtain a coefficient whose square is more than three times the expectancy, we can state that the probability that it is produced by a chance distribution of the quantities analysed is $\frac{1}{20}$. If the square of the Fourier coefficient be 12 times its expectancy, the probability that 
it is produced by a chance distribution is 1 in 160,000. 
\end{quotation}

But, as noted in \S\ref{subsec:acknowledge}, Brunt set 19 to 1 as the odds for practical certainty in his periodogram analysis of Greenwich temperature records in 1919
\cite[p.~328]{Brunt:1919}.

On pp.~131--132 of his book, Brunt summarized the controversy concerning the rejection of observations as follows:
\begin{quotation}
\noindent
The whole question of the possibility of rejecting observations on the ground of theoretical discussion based on residuals only, has given rise to a considerable amount of controversy. Bessel opposed the rejection of any observation unless the observer was satisfied that the external conditions produced some unusual source of error not present in the other observations of the series. Peirce's criterion was at an early date subjected to very severe criticism. Airy claimed that it was defective in foundation, and illusive in its results. He maintained that, so long as the observer was satisfied that the same sources of error were at work, though in varying degrees, throughout a series of observations, the computer should have no right to reject any observation by a discussion based solely on antecedent probability.  An observation should be rejected only when a thorough examination showed that the causes of error normally at work were not sufficient to produce the error in the doubtful observation. Airy also cited a case where the rejection of the observations having large residuals led to poor results. \ldots

Though many of the arguments of Airy and others against the use of mathematical criteria such as Peirce's have been shown to be based on faulty premises, the fact remains that none of these criteria have ever come into general use.
\end{quotation}
Brunt concludes, however, by subscribing to a simple rule for ``a moderate number of observations'':  reject observations for which the residual exceeds five probable errors and take a close look at any others whose residual exceeds 3.5 probable errors.  This may have been influential; in 1931, Harold Jeffreys reported that rejecting observations with residuals greater five probable errors was ``a common astronomical practice'' \cite[p.~83]{Jeffreys:1931}.

\subsection{Ronald A.\ Fisher, 1890--1962}\label{subsec:Fisher}

Fisher is celebrated as the most accomplished mathematical statistician of the 20th century.  He laid out his understanding of ``tests of significance'', by no means his most important contribution, in his monograph \emph{Statistical Methods for Research Workers} \cite{Fisher:1925}, published in 1925 and in many subsequent editions.

\subsubsection*{Shelving the probable error}

As we saw in \S\ref{subsec:Fisher1}, the most novel aspect of the 1925 book was that it tabulated values of the tail probability P not only for the normal distribution and Pearson's $\chi^2$ but also for a number of other distributions that can be used when the assumption that individual observations have a normal distribution is taken seriously, including Student's $t$ and the distribution of the correlation coefficient.  As Fisher explains in the following passage, this led him  to abandon measurement in terms of the probable error in favor of measurement in terms of tail probabilities, and in particular to replace the criterion of two probable errors by the criterion of 5\%.
\begin{quote}
Pp. 47--48:  ``The value of the deviation beyond which half the observations lie is called the quartile distance, and bears to the standard deviation the ratio $\cdot67449$.  It is therefore a common practice to calculate the standard error and then, multiplying it by this factor, to obtain the probable error. The probable error is thus about two-thirds of the standard error, and as a test of significance a deviation of three times the probable error is effectively equivalent to one of twice the standard error. The common use of the probable error is its only recommendation; when any critical test is required the deviation must be expressed in terms of the standard error in using the probability integral table.''
\end{quote}

\subsubsection*{The end of Edgeworthian signifying}

Here are some examples the book's non-Edgeworthian use of ``significant''.  
\begin{itemize}
\item
P. 21:  ``The table illustrates the general fact that the significance in the normal distribution of deviations exceeding four times the standard deviation is extremely pronounced.''
\item
P. 123:  ``This suggests the possibility that if we had fitted a more complex regression line to the data the probable errors would be further reduced to an extent which would put the significance of b beyond doubt.''
\item
Pp. 158--159: ``Taking the four definite levels of significance, represented by $\text{P} = \cdot10$, $\cdot05$, $\cdot02$, and $\cdot01$, the table shows for each value of $n$, from 1 to 20, and thence by larger intervals to 100, the corresponding values of $r$.''
\item
P. 90:  ``the significance will become more and more pronounced as the sample is increased in size\ldots''
\item
P. 47, with reference to the table for the normal distribution:  ``The value for which $\text{P} = \cdot05$, or 1 in 20, is $1\cdot96$ or nearly 2; it is convenient to take this point as a limit in judging whether a deviation is to be considered significant or not. Deviations exceeding twice the standard deviation are thus formally regarded as significant. Using this criterion, we should be led to follow up a false indication only once in 22 trials, even if the statistics were the only guide available.  
\item
Pp. 81--82: ``The expected values are calculated from the observed total, so that the four classes must agree in their sum, and if three classes are filled in arbitrarily the fourth is therefore determinate, hence $n=3$, $\chi^2 =10.87$, the chance of exceeding which value is between .01 and .02; if we take $\text{P}=.05$ as the limit of significant deviation, we shall say that in this case the deviations from expectation are clearly significant.''
\item
Pp. 102--102:  ``If, therefore, we know the standard deviation of a population, we can calculate the standard deviation of the mean of a random sample of any size, and so test whether or not it differs significantly from any fixed value. If the difference is many times greater than the standard error, it is certainly significant, and it is a convenient convention to take twice the standard error as the limit of significance; this is roughly equivalent to the corresponding limit $\text{P} = \cdot05$, already used for the $\chi^2$ distribution.''
\item
P. 158:  ``very much exaggerating the significance.''
\item
P. 161:  ``The values given in Table V. (A) for $n = 25$, and $n = 30$, give a sufficient indication of the level of significance attained by this observation.
\end{itemize}
It is also notable that we find the term ``statistical significance'' (page 218).

\paragraph{Preface to the 6th edition (1936).}

As John Aldrich has pointed out to me, Fisher discusses the nature of his book and touches on the issue of p-hacking in the preface to the 6th edition of \emph{Statistical Methods for Research Workers}, published in 1936.  We find this argument on pp.~ix--x:
\begin{quotation}
Those critics who would like to have seen the inclusion of mathematical proofs of the more important propositions of the underlying theory, must still be referred to the technical publications given in the list of sources. There they will encounter exactly those difficulties which it would be undesirable to import into the present work; and will perceive that modern statistics could not have been developed without the elaboration of a system of ideas, logical and mathematical, which, however fascinating in themselves, cannot be regarded as a necessary part of the equipment of every research worker. 

To present ``elementary proofs,'' of the kind which do not involve these ideas, would be really to justify the censure of a second school of critics, who, rightly feeling that a fallacious proof is worse than none, are eager to decry any attempt to ``teach people to run before they can walk.'' The actual scope of  the present volume really exempts it from this criticism, which, besides, in an age of technical co-operation, has seldom much force. The practical application of general theorems is a different art from their establishment by mathematical proof. It requires fully as deep an understanding of their meaning, and is, moreover, useful to many to whom the other is unnecessary.  
\end{quotation}
And then we find this paragraph on p.~xii:
\begin{quotation}
I am indebted to Dr W.\ E.\ Deming for the extension of the table of $z$ to the $0.1$ per cent.\ level of significance.  Such high levels of significance are especially useful when the test we make is the most favourable out of a number which \emph{a priori} might equally well have been chosen.  Colcord and L.\ S.\ Deming have published a slightly fuller Table in the \emph{Indian Journal of Statistics} (1936).
\end{quotation}

\subsection{Jerzy Neyman, 1894--1981}\label{subsec:Neyman}

As mentioned in \S\ref{subsec:returnpc}, Cournot defined what we now call a confidence interval in his 1843 \emph{Exposition}, and such intervals were widely used in the 19th century.  Yet the notion of a confidence interval is now widely attributed to Jerzy Neyman.  How did this come about?

Neyman introduced the terms \emph{confidence coefficient} and \emph{confidence interval} in English, in 1934 \cite{Neyman:1934}.  Already, beginning in 1930, Fisher had published examples of probability intervals for parameters obtained without the use of prior probabilities.  Fisher called his intervals \emph{fiducial}, and few of his readers, then or now, have been able to find a discern a consistent theory behind them.  Confronted with this reality and Fisher's stature, Neyman tried to credit Fisher with priority while presenting his own definition as a clarification; he explained that Fisher's papers\ldots
\begin{quote}
\ldots have been misunderstood and the validity of statements they contain formally questioned. This I think is due largely to the very condensed form of explaining ideas used by R. A. Fisher, and perhaps also to a somewhat difficult method of attacking the problem.
\end{quote}
He also credited Markov with having considered, in the context of least squares, the problem of finding the narrowest confidence intervals for a given confidence coefficient.  This is a evidently a reference to Gauss's theorem, according to which the arithmetic mean has the smallest variance among linear unbiased estimators.  We may conjecture that Neyman learned the erroneous attribution to Markov from Bernstein; as a result, the theorem is now called the \emph{Gauss-Markov theorem}. 

Beyond granting credit to Fisher and Markov, Neyman also pointed out that confidence intervals were already widely used in practice.  He wrote (p.~562),
\begin{quote}
\ldots the methods of estimating, particularly in the case of large samples, resulting from the work of Fisher, are often precisely the same as those which are already in common use. Thus the new solution of the problems of estimation consists mainly in a rigorous justification of what has been generally considered correct more or less on intuitive grounds.
\end{quote}
Neyman was evidently unaware that Laplace, Cournot, and Bienaym\'e had explained the large-sample agreement between the inverse and direct methods for obtaining probability intervals.  Perhaps he was still unaware of this 19th century work in the 1950s, but by that time he had discovered the theorem in an article by von Mises in 1919 \cite{von-Mises:1919b} and had also discerned it, perhaps between the lines, in Bernstein's 1927 book.  He and his student Lucien Le Cam then dubbed the theorem the \emph{Bernstein-von Mises theorem} \cite{Le-Cam:1956}.  

Fisher was not happy with Neyman's appreciation of his work, and Neyman was forced to present his theory as different from Fisher's; see \cite[Chapter 6]{Lehmann:2011}.   In 1941 \cite{Neyman:1941}, Neyman explained that he had in fact been using confidence intervals in Poland beginning in 1930, before he had known about Fisher's work.  His Polish name for them had been  \emph{przedzia\l\ ufno\'sci}.  Both the English \emph{confidence} and the Polish \emph{ufno\'s\'c} refer to the attitude of a person who is sure of something.  Neyman's teacher Bernstein may also have had some influence here; \emph{ufno\'s\'c} is a reasonable translation into Polish of Bernstein's Russian \begin{cyr}уверенность\end{cyr}.
 
In 1937, Neyman encountered difficulties when he tried to publish a more extensive theoretical study of confidence intervals in the \emph{Journal of the Royal Statistical Society}.  As he explained many years later to his biographer Constance Reid \cite[p.~139]{Reid:1998}, one referee (later known to be Yule, who could scarcely have seen anything novel in the notion of a confidence interval) was unfavorable, the other (later known to be A.\ C.\ Aitken) was favorable.  Aitken advised the editor that the paper might benefit from using the foundation for mathematical probability recently published by Kolmogorov \cite{Kolmogorov:1933}.  Neyman had not previously known about Kolmogorov's measure-theoretic axiomatization for probability, but he made it central to a successful revision \cite{Neyman:1937}.  Now there was prestigious mathematical backing for Neyman's contention that he was providing a previously absent theoretical basis for an established practice.   He made the point this way, on pp. 346--347 of the 1937 article:
\begin{quotation}
If we look through a number of recent statistical publications, we shall find that it is exceedingly rare that the values of unique estimates are given without the $\pm \text{S}_\text{T}$.  We shall find also that the comments on the values of $\text{T}$ are largely dependent on those of $\text{S}_\text{T}$. This shows that what the statisticians have really in mind in problems of estimation is not the idea of a unique estimate but that of two estimates having the form, say
\[
      \underline{\theta} - k_1 \text{S}_\text{T} \qquad \text{and} \qquad \overline{\theta} + k_2 \text{S}_\text{T},
\]
where $k_1$ and $k_2$ are certain constants, indicating the limits between which the true value of $\theta$ presumably falls.

In this way the practical work, which is frequently in advance of the theory, brings us to consider the theoretical problem of estimating the parameter $\theta$ by means of the interval $(\underline{\theta}, \overline{\theta})$, extending from $\underline{\theta}$ to $\overline{\theta}$. These limits will be called the lower and upper estimates of $\theta$ respectively. \dots
\end{quotation}

When Russian mathematicians later discussed Fisher's work, they translated \emph{fiducial limits} as \begin{cyr}доверительные границы\end{cyr}; like \emph{fiducial}, the adjective \begin{cyr}доверитель\-ный\end{cyr} evokes the notion of trust.  Bernstein rejected Fisher's reasoning in favor of the classical (Bayesian) argument \cite{Bernstein:1941}, whereas Kolmogorov favored Neyman's theory \cite{Kolmogorov:1942}.  Kolmogorov's view carried the day, but the Russians retained the Fisherian adjective; a confidence interval is now a \begin{cyr}доверительный интервал\end{cyr} in Russian.  Perhaps \emph{confidence} could not be translated back into \begin{cyr}уверенность\end{cyr}, because that word was already taken by the classical theory; see~\S\ref{subsec:Bernstein}.

\subsection{Egon S. Pearson, 1895--1980}\label{subsec:Pearson-ES}

When Pearson, Karl Pearson's son, joined his father's department at the University of London in 1921, he had already encountered the theory of errors and statistical methods at Cambridge, where he had interacted with Eddington and Yule.  Passing comments in his early articles suggest that by the early 1920s he and his fellow British statisticians were all too aware that p-values are invalidated when tests are selected on the basis of data.

In 1925, for example, we find the following comments in an article in which Pearson tried to evaluate Bayes's theorem empirically \cite[p.~435]{Pearson-ES:1925}:
\begin{quotation}
\noindent
\ldots If, to take a different example, in our statistical experience, we only apply the $\chi^2$ Test for Goodness of Fit to cases where from inspection the fit looks unsatisfactory, then clearly we must be very careful what deductions we draw. For we might obtain in nearly 100\% of tests, fits with $P(\chi^2)$ less than $0\cdot1$, where yet in every case the sample had been drawn from the supposed population.
 
It is of course the old difficulty; once or twice in a thousand times an event will occur against which the odds are 999 to 1, but when it does occur we are inclined to think that something is wrong, focusing our attention on the single event and forgetting the 999 times when we have observed that it did not happen.
\end{quotation}

A decade later, in 1936, we find this passage in Pearson's article with Chandra Sekar on the rejection of outliers \cite[p.~317]{Pearson/ChandraSekar:1936}:
\begin{quote}
\ldots In conclusion, since it is sometimes held that the appropriate test can be chosen after examining the data in the sample, a final word of caution is necessary. To base the choice of the test of a statistical hypothesis upon an inspection of the observations is a dangerous practice; a study of the configuration of a sample is almost certain to reveal some feature, or features, which are exceptional if the hypothesis is true.
\end{quote}

\subsection{Morris Viteles, 1898--1996}\label{subsec:Viteles}

In a brief article on intelligence testing published by Viteles in 1922 \cite{Viteles:1922}, we find ``greatly significant'', ``particularly significant'', and ``high enough to be of considerable significance''.  We also see the first use of ``level of significance'' that I have found.  Viteles states
\begin{quote}
\ldots reduces the co-efficient of correlation \ldots to plus $.21\pm.091$, much below the level of significance.
\end{quote}
and
\begin{quote}
\ldots reduces the co-efficient of correlation of these two tests to plus $0.37\pm.080$, also below the level of significance.
\end{quote}
Here the level of significance is evidently six probable errors.  Viteles, who spent most of his career at the University of Pennsylvania, had not benefited from a year with Pearson, but he became a prominent figure in industrial and organizational psychology.

\section{Bernoullian and Bayesian}\label{app:bernoullian}

On p.~\pageref{p:bernoullian}, I cited some authors who have used the adjective \emph{Bernoullian} rather than \emph{frequentist} to designate statistical methods that follow Jacob Bernoulli's example rather than that of Thomas Bayes.  Here are some quotations from those authors.
\begin{itemize}
\item
Francis Edgeworth used \emph{Bernoullian} in this way in 1918, contrasting ``the \emph{direct} problem associated with the name of Bernoulli'' with ``the \emph{inverse} problem associated with the name of Bayes'' \cite{Edgeworth:1918}.  
\item
Richard von Mises made a similar remark in German in 1919 (\cite{vonMises:1919}, page 5):  ``Man kann die beiden großen Problemgruppen \ldots als den Bernoullischen und den Bayesschen Ideenkreis charakterisieren.''  In English:  ``We can call the two large groups of problems the  Bernoullian and Bayesian circles of ideas.'' 
\item
A.\ P.\ Dempster advocated the usage in 1966 \cite{Dempster:1966}.  In 1968 \cite{Dempster:1968d}, in a review of three volumes of collected papers by Jerzy Neyman and E.\ S.\ Pearson, Dempster wrote
\begin{quote}
Neyman and Pearson rode roughshod over the elaborate but shaky logical structure of Fisher, and started a movement which pushed the Bernoullian approach to a high-water mark from which, I believe, it is now returning to a more normal equilibrium with the Bayesian view.
\end{quote}
\item
Ian Hacking used \emph{Bernoullian} repeatedly in his 1990 book, \emph{The Taming of Chance} \cite{Hacking:1990}.  Writing about Poisson's interest in changes in the chance of conviction by a jury, he wrote (page 97):
\begin{quote}
Laplace had two ways in which to address such questions.  One is Bernoullian, and attends to relative frequencies; the other is Bayesian, and is usually now interpreted in terms of degrees of belief.  Laplace almost invited his readers not to notice the difference.
\end{quote}
\end{itemize}
This usage would recognize Bernoulli as the first to state a theory of direct statistical estimation, just as Bayes was the first to state Bayes's formula.  It would also allow us to contrast Bernoullian and Bayesian methods without asserting anything about how probabilities are to be interpreted.  Using \emph{frequentist} for both an interpretation of probability and a method of inference is a source of conceptual and historical confusion.  It obscures, for example, the fact that von Mises, long recognized as the leading proponent of ``the frequency theory of probability'', always contended that Bayes's formula provides the correct method of statistical inference \cite{vonMises:1942}.

\end{appendices}

\addcontentsline{toc}{section}{References}
\bibliographystyle{plain}
\bibliography{55ref}

\end{document}